\documentclass[12pt]{iopart}

\usepackage{iopams}  
\usepackage{color}
\usepackage{graphicx}
\begin{document}

\title{Coupling of RF Antennas to Large Volume Helicon Plasma}

\author{Lei Chang$^{1, 3}$, Xinyue Hu$^1$, Lei Gao$^1$, Wei Chen$^1$, Xianming Wu$^2$, Xinfeng Sun$^2$, Ning Hu$^3$, and Chongxiang Huang$^1$}
\address{$^1$ School of Aeronautics and Astronautics, Sichuan University, Chengdu 610065, China}
\address{$^2$ National Key Laboratory of Science and Technology on Vacuum Technology and Physics, Lanzhou Institute of Physics, Lanzhou 730000, China}
\address{$^3$ College of Aerospace Engineering, Chongqing University, Chongqing 400044, China}
\ead{\textcolor{blue}{leichang@scu.edu.cn}, chxhuang@scu.edu.cn}

\date{\today}

\begin{abstract}
Large volume helicon plasma sources are of particular interest for large scale semiconductor processing, high power plasma propulsion and recently plasma-material interaction under fusion conditions. This work is devoted to studying the coupling of four typical RF antennas to helicon plasma with infinite length and diameter of $0.5$~m, and exploring its frequency dependence in the range of $13.56-70$~MHz for coupling optimization. It is found that loop antenna is more efficient than half helix, Boswell and Nagoya III antennas for power absorption; radially parabolic density profile overwhelms Gaussian density profile in terms of antenna coupling for low-density plasma, but the superiority reverses for high-density plasma. Increasing the driving frequency results in power absorption more near plasma edge, but the overall power absorption increases with frequency. Perpendicular stream plots of wave magnetic field, wave electric field and perturbed current are also presented. This work can serve as an important reference for the experimental design of large volume helicon plasma source with high RF power. 
\end{abstract}



\maketitle

\section{Introduction}\label{int}
Large volume helicon plasma has been attracting growing interest in various fields, including large scale semiconductor processing\cite{Lieberman:2005aa}, electrodeless plasma propulsion with high power input\cite{Diaz:2000aa, Diaz:2001aa, Shinohara:2013aa, Shinohara:2014aa}, and emerging plasma-material interaction under fusion conditions recently\cite{Rapp:2016aa, Rapp:2017aa, Goulding:2017aa, Blackwell:2012aa}. Large volume allows high power capacity and big heat flux, besides large cross sectional area. Except the novel design of spiral antenna to excite helicon plasma from an axial end and with diameter of $75$~cm\cite{Shinohara:2004aa, Tanikawa:2006aa}, most helicon sources employ cylindrical RF (Radio Frequency) antennas wrapping plasma with diameter less than $40$~cm. For extremely high power applications such as MPEX (Material Plasma Exposure eXperiment)\cite{Rapp:2016aa, Rapp:2017aa, Goulding:2017aa}, the usually involved quartz tube is even removed. Although half helix, Boswell, Nagoya III and loop antennas are widely used for small helicon sources, it is still not yet clear that which type of RF antenna can best couple helicon plasma of large size. The presented work is devoted to comparing these typical helicon antennas in terms of power absorption and wave field structure. The motivation is to assist experimental design of RF antenna for helicon sources with large diameter. Since high power is not the particular focus of this work, quartz tube is still kept to separate plasma and antenna radially. A C++ program based on Maxwell's equations and cold-plasma dielectric tensor will be employed to compute the spatial power absorption and wave field for different antennas, and their comparisons will be analyzed for experimental reference. 

\section{Theoretical model and computational scheme}\label{mdl}
To study the coupling of RF antennas to large volume helicon plasma, we utilize the well-known $\textbf{HELIC}$ code developed by D. Arnush and F. F. Chen\cite{Chen:1997aa, Arnush:1997aa, Arnush:2000aa}. The underlying theoretical model consists of Maxwell's equations for a radially nonuniform plasma with standard cold-plasma dielectric tensor. These equations can be manipulated to give the following set of coupled differential equations for the Fourier transformed variables: 
\begin{equation}
\frac{\partial E_\varphi}{\partial r}=\frac{i m}{r} E_r-\frac{E_\varphi}{r}+i\omega B_z,
\end{equation}
\begin{equation}
\frac{\partial E_z}{\partial r}=i k E_r-i\omega B_\varphi,
\end{equation}
\begin{equation}
i\frac{\partial B_\varphi}{\partial r}=\frac{m}{r}\frac{k}{\omega}E_\varphi-\frac{i B_\varphi}{r}+\left(P-\frac{m^2}{k_0^2 r^2}\right)\frac{\omega}{c^2}E_z,
\end{equation}
\begin{equation}
i\frac{\partial B_z}{\partial r}=-\frac{\omega}{c^2}i D E_r+\left(k^2-k_0^2 S\right)\frac{E_\varphi}{\omega}+\frac{m}{r}\frac{k}{\omega}E_z.
\end{equation}
Here, a cylindrical coordinate system $(r, \varphi, z)$ and first-order perturbation form of $\exp[i(m \varphi+k z-\omega t)]$ have been chosen, with $m$ azimuthal mode number, $k$ axial mode number and $\omega$ wave frequency. The variables $E$ and $B$ are wave electric field and wave magnetic field, respectively, and $k_0=\omega/c$ with $c$ the speed of light. The cold-plasma dielectric elements $S$, $D$ and $P$ are expressed below following the notation of Stix\cite{Stix:1957aa, Stix:1992aa}: 
\textcolor{blue}{
\begin{equation}
R=1-\sum_s\frac{\omega_{ps}^2}{\omega(\omega+i\nu_s)}\left[\frac{(\omega+i\nu_s)}{(\omega+i\nu_s)+\omega_{cs}}\right],
\end{equation}
\begin{equation}
L=1-\sum_s\frac{\omega_{ps}^2}{\omega(\omega+i\nu_s)}\left[\frac{(\omega+i\nu_s)}{(\omega+i\nu_s)-\omega_{cs}}\right],
\end{equation}
\begin{equation}
S=\frac{1}{2}(R+L),~D=\frac{1}{2}(R-L),~P=1-\sum_s\frac{\omega_{ps}^2}{\omega(\omega+i\nu_s)},
\end{equation}
where a phenomenological collision rate $\nu$ has been introduced to resolve electron-neutral and electron-ion collisions.} Here $\omega_{ps}=\sqrt{q_s^2n_{s}/(\epsilon_0m_s)}$ is the plasma frequency, and $\omega_{cs}=q_sB_0/m_s$ is the cyclotron frequency, with $q_s$ electric charge, $m_s$ particle mass and $n_s$ plasma density. The subscript $s$ denotes the plasma species, namely electron and ions, and $0$ denotes the equilibrium value. The parameters $R$ and $L$ stands for right-hand and left-hand circularly polarized modes, respectively. These expressions are perfectly general and capable of including multiple ion species, together with displacement current. We consider a single ion species with single charge in this work. 

\begin{figure}[ht]
\begin{center}$
\begin{array}{ll}
\includegraphics[width=0.46\textwidth,angle=0]{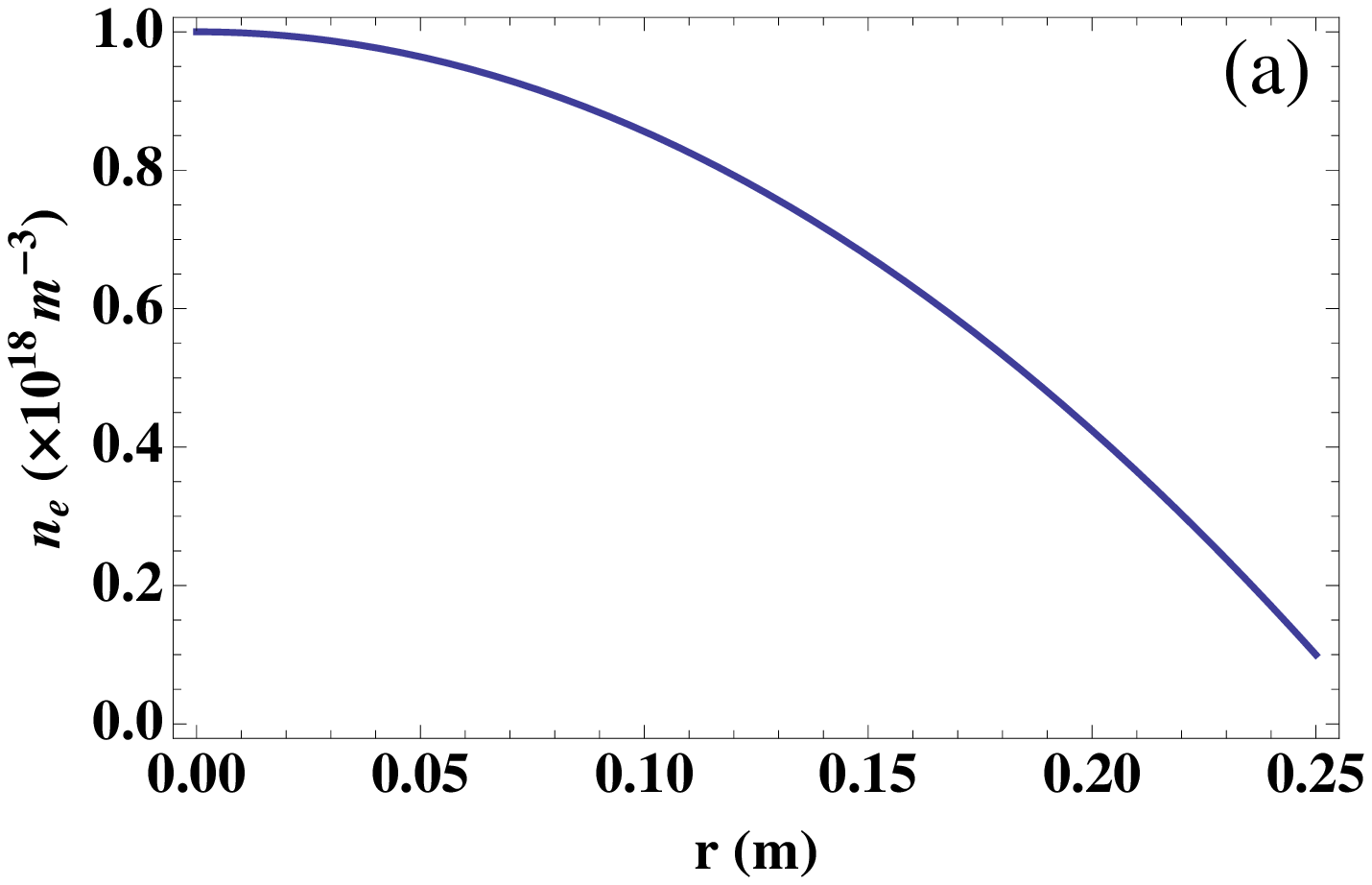}&\includegraphics[width=0.463\textwidth,angle=0]{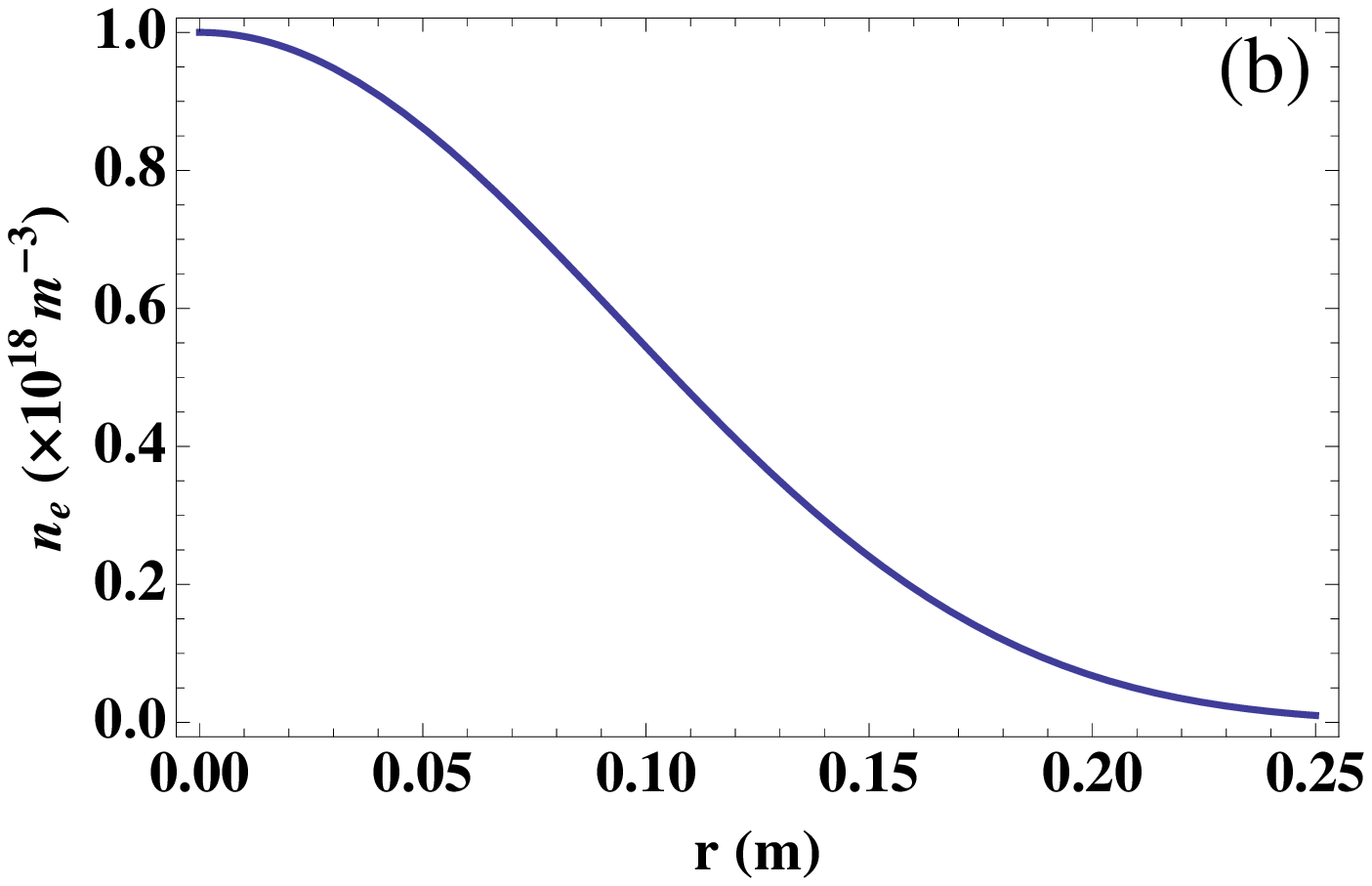}
\end{array}$
\end{center}
\caption{Radial profiles of plasma density: (a) parabolic, (b) Gaussian.}
\label{fg1}
\end{figure}

The code solves Eqs.~(1-4) for each value of $k$ through a standard subroutine, and $m$ is fixed for each calculation. To construct a large volume helicon plasma source, we set the radii of plasma and antenna to be $0.25$~m and $0.255$~m, respectively, and expand the enclosing cavity to be $0.35$~m in radius. The length is assumed to be infinite, although the axial range of $-0.63<z<0.63$~m is computed. The cavity is metal, leading to conducting boundary conditions. Referring to most helicon discharges\cite{Boswell:1987aa, Blackwell:2012aa, Squire:2006aa, Mori:2004aa}, we choose argon and spectral range of $5<k<50~\mbox{m}^{-1}$, and set the background pressure and temperature to be $3$~eV and $10$~mTorr, respectively. Moreover, we consider four types of RF antennas: half helix, Boswell and Nagoya III to excite $m=1$ mode and loop to excite $m=0$ mode. Their length is the same, namely $0.2$~m. Two plasma density profiles which are commonly measured in experiments are employed: parabolic and Gaussian in radius. As shown in Fig.~\ref{fg1}, these two density profiles have the same peak value on axis but different gradients in radius. Their edge values are $0.1$ and $0.01$ respectively of the peak value. \textcolor{blue}{As pointed earlier, the radial density gradient and accordingly edge density have an important effect on the non-resonant mode conversion of helicon wave into Trivelpiece-Gould wave and power absorption, we thereby choose these two density profiles with very different radial gradients.\cite{Shamrai:1998aa}} The density level of $10^{18}~\mbox{m}^{-3}$ is also typical for helicon discharges\cite{Boswell:1970aa, Boswell:1984ab, Boswell:1997aa, Chen:1997ab, Chang:2011aa}. All computed results are normalized to antenna current of $1$~A, therefore, the presented power absorptions are relative results. \textcolor{blue}{The background magnetic field strength is $0.02$~T.}

\section{Typical RF antennas}\label{ant}
\begin{figure}[ht]
\begin{center}$
\begin{array}{ll}
\includegraphics[width=0.48\textwidth,angle=0]{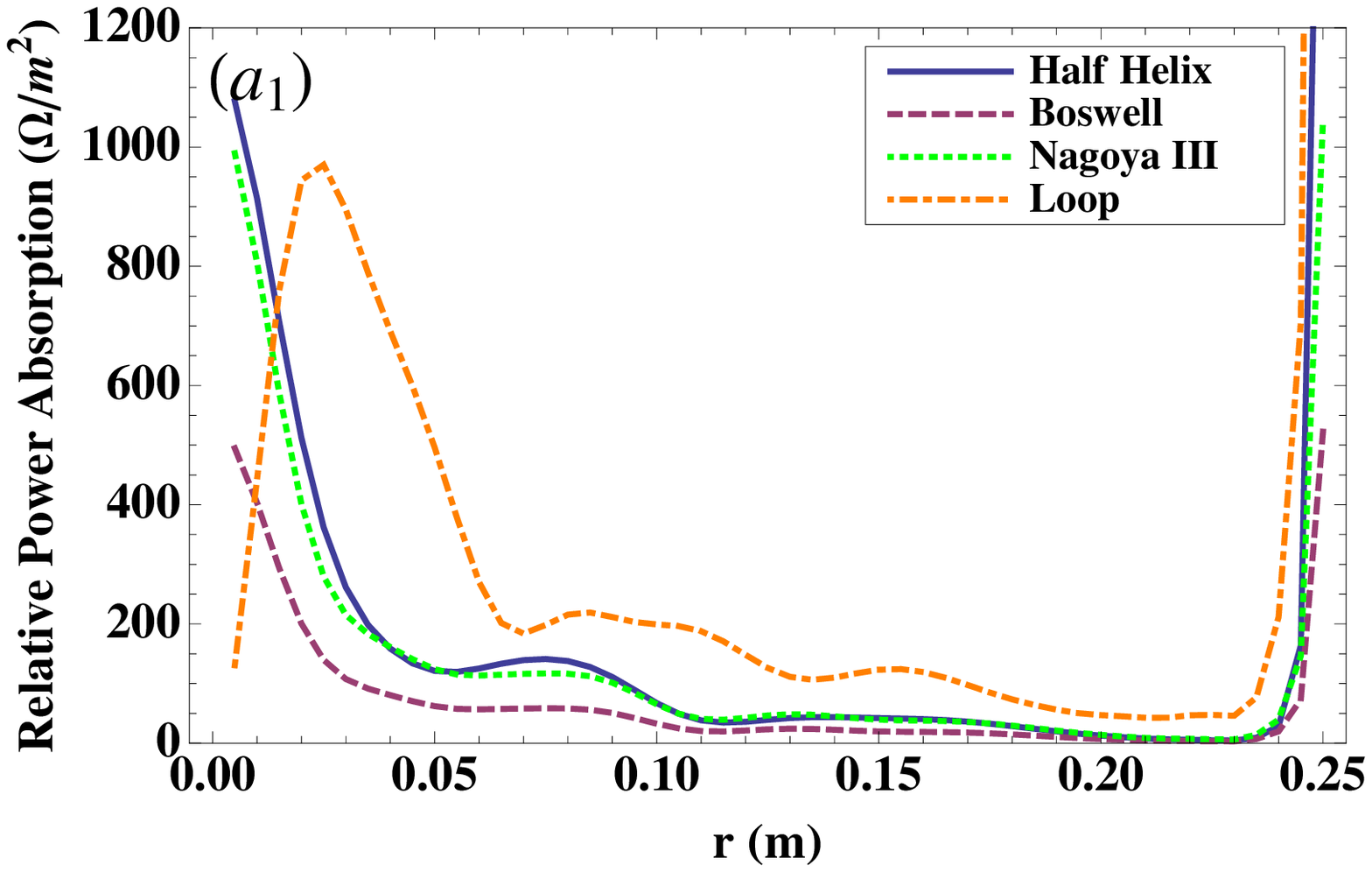}&\includegraphics[width=0.471\textwidth,angle=0]{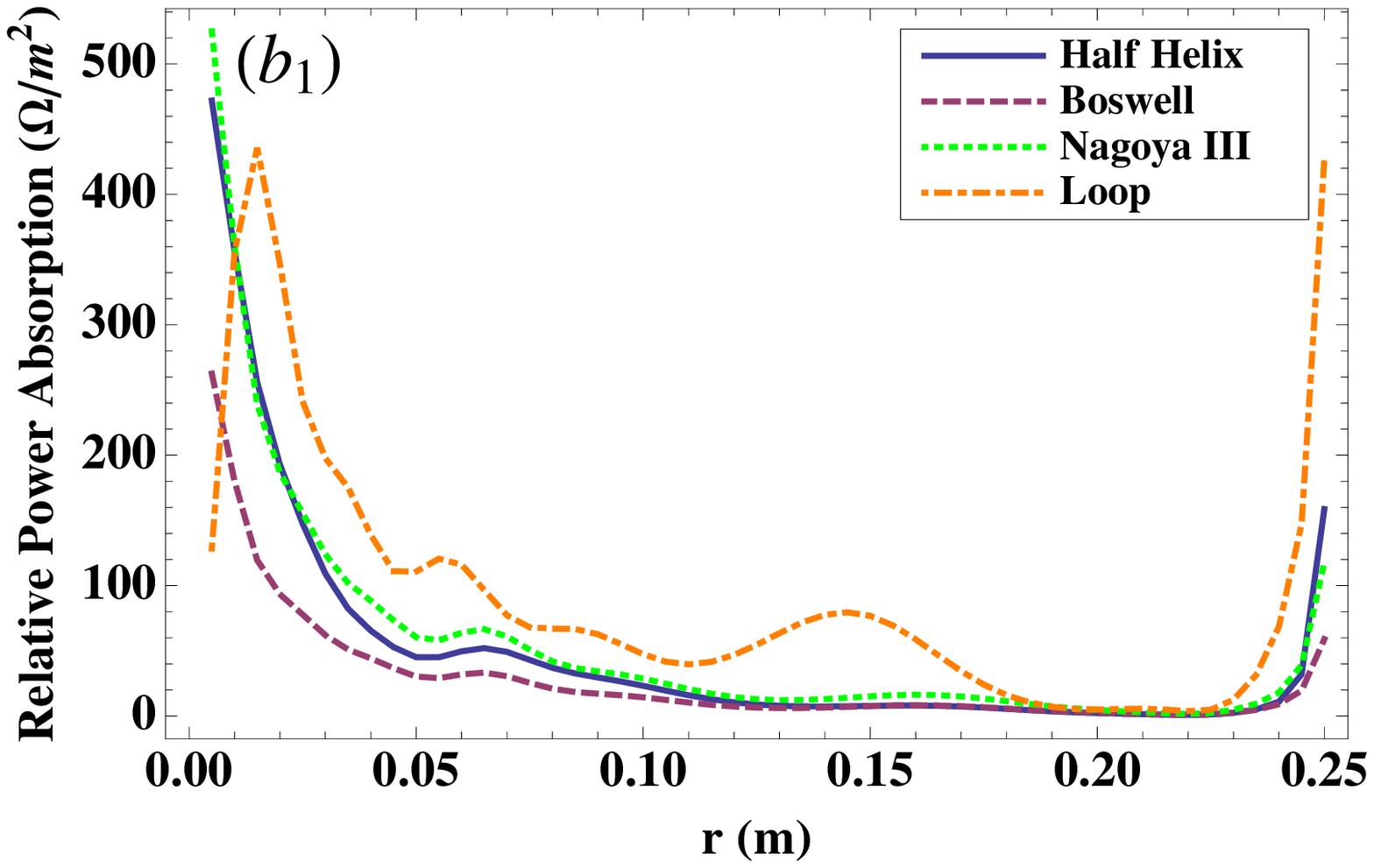}\\
\hspace{0.35cm}\includegraphics[width=0.465\textwidth,angle=0]{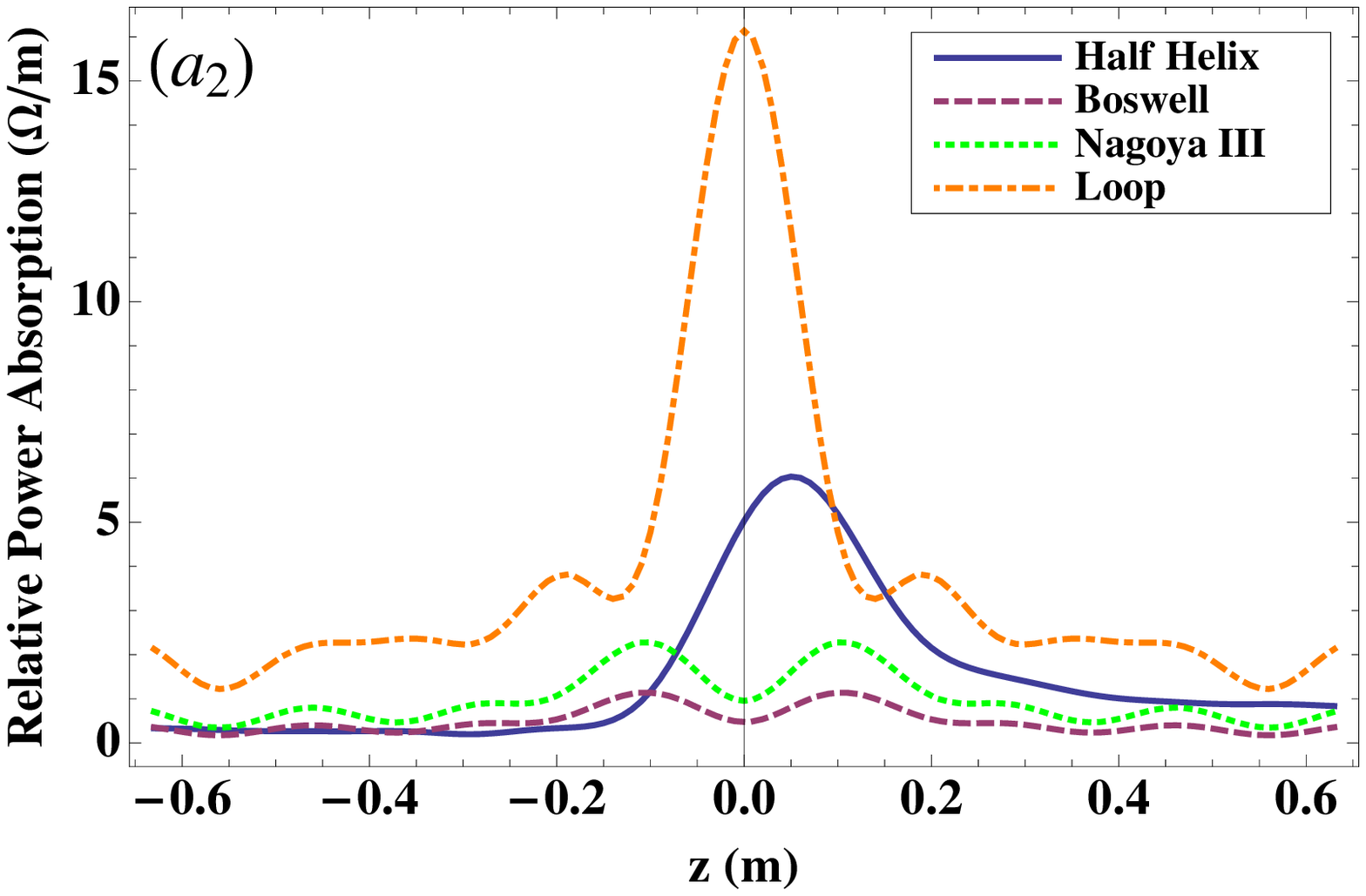}&\hspace{0.13cm}\includegraphics[width=0.471\textwidth,angle=0]{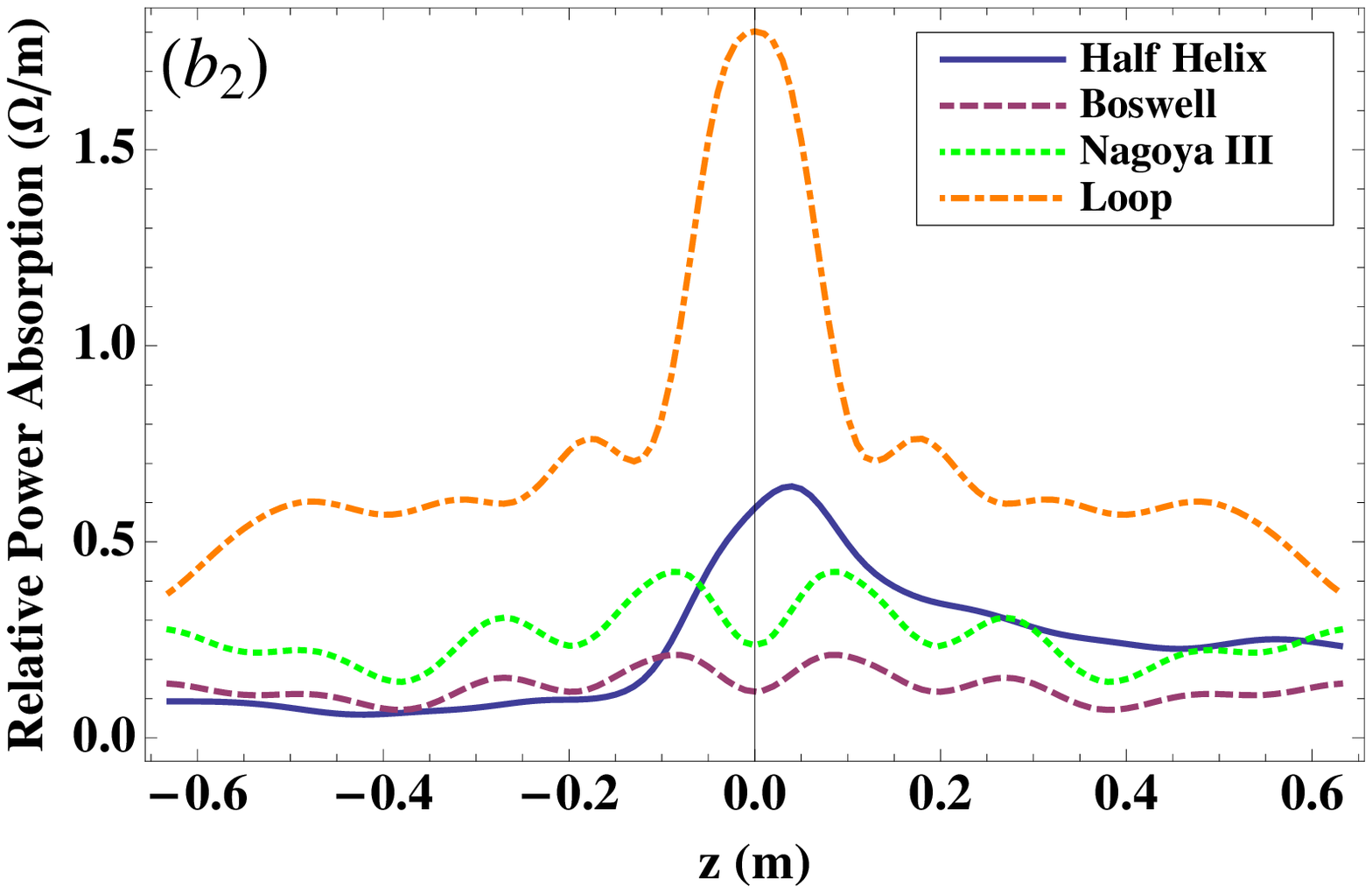}
\end{array}$
\end{center}
\caption{Relative power absorption in radial ($z=0.2$~m) and axial ($r=0.02$~m) directions for four typical RF antennas: (a) parabolic density profile, (b) Gaussian density profile.}
\label{fg2}
\end{figure}

\begin{figure}[ht]
\begin{center}$
\begin{array}{ccc}
\includegraphics[width=0.3\textwidth,angle=0]{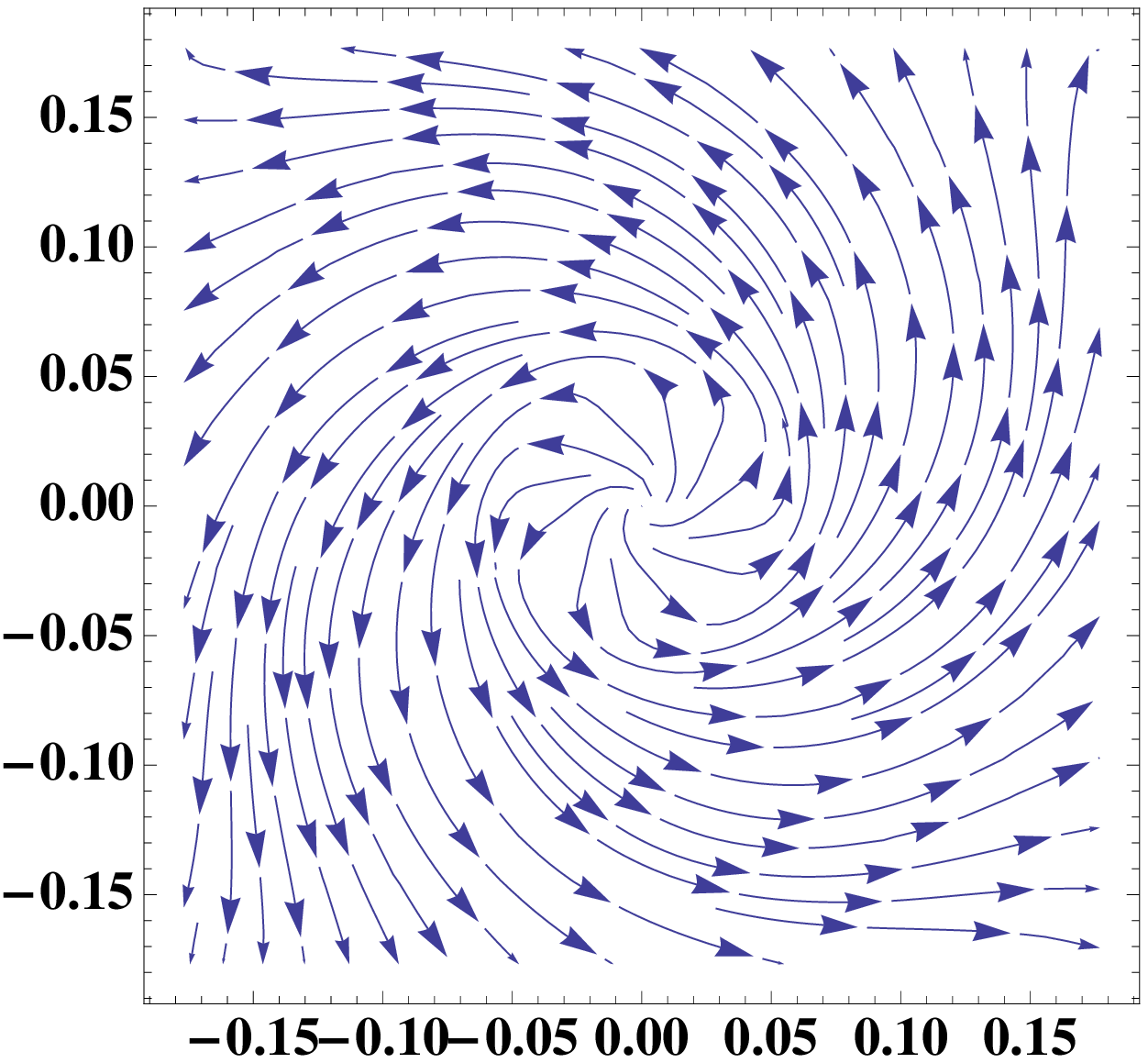}&\hspace{-0.35cm}\includegraphics[width=0.3\textwidth,angle=0]{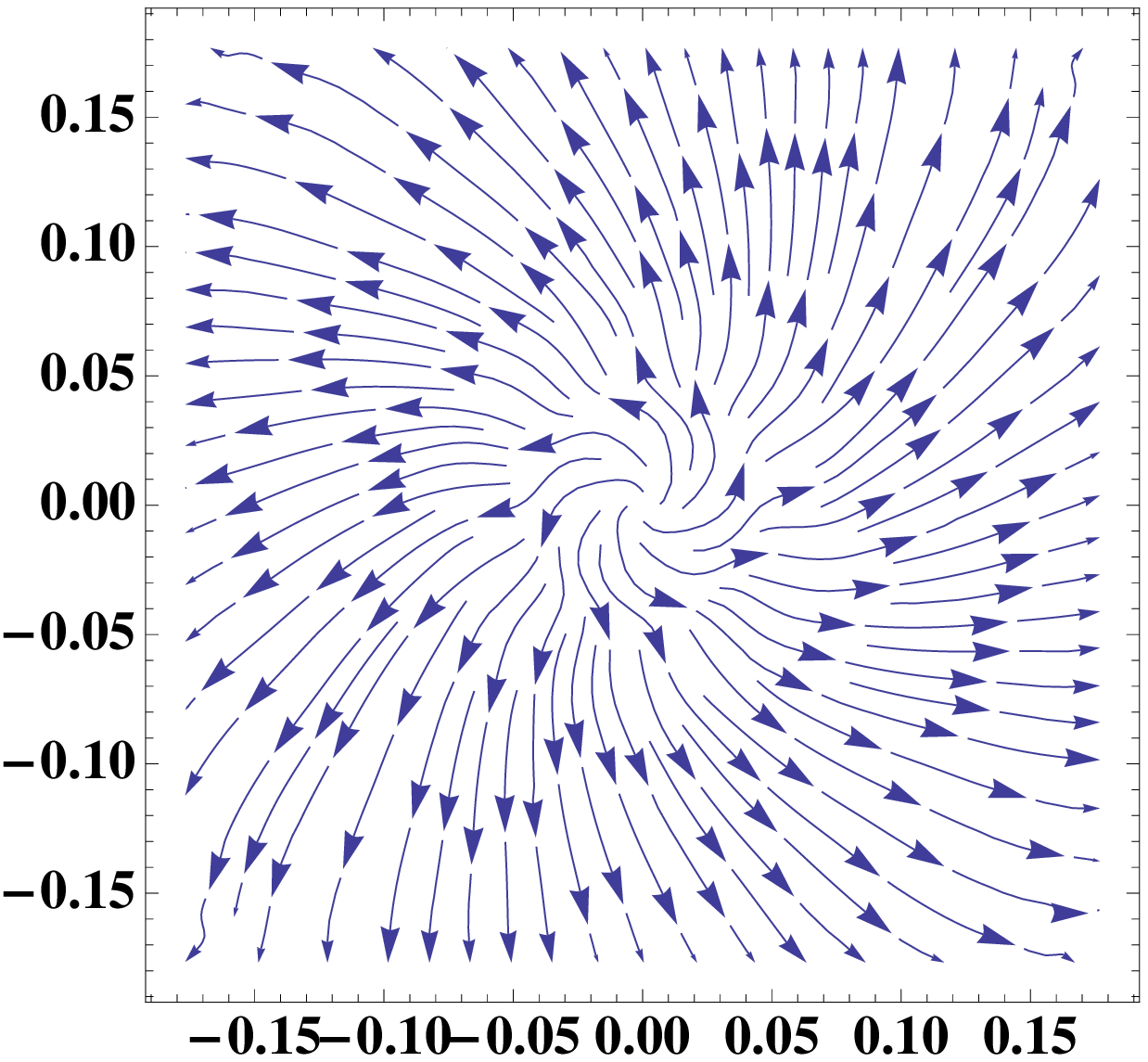}&\hspace{-0.35cm}\includegraphics[width=0.3\textwidth,angle=0]{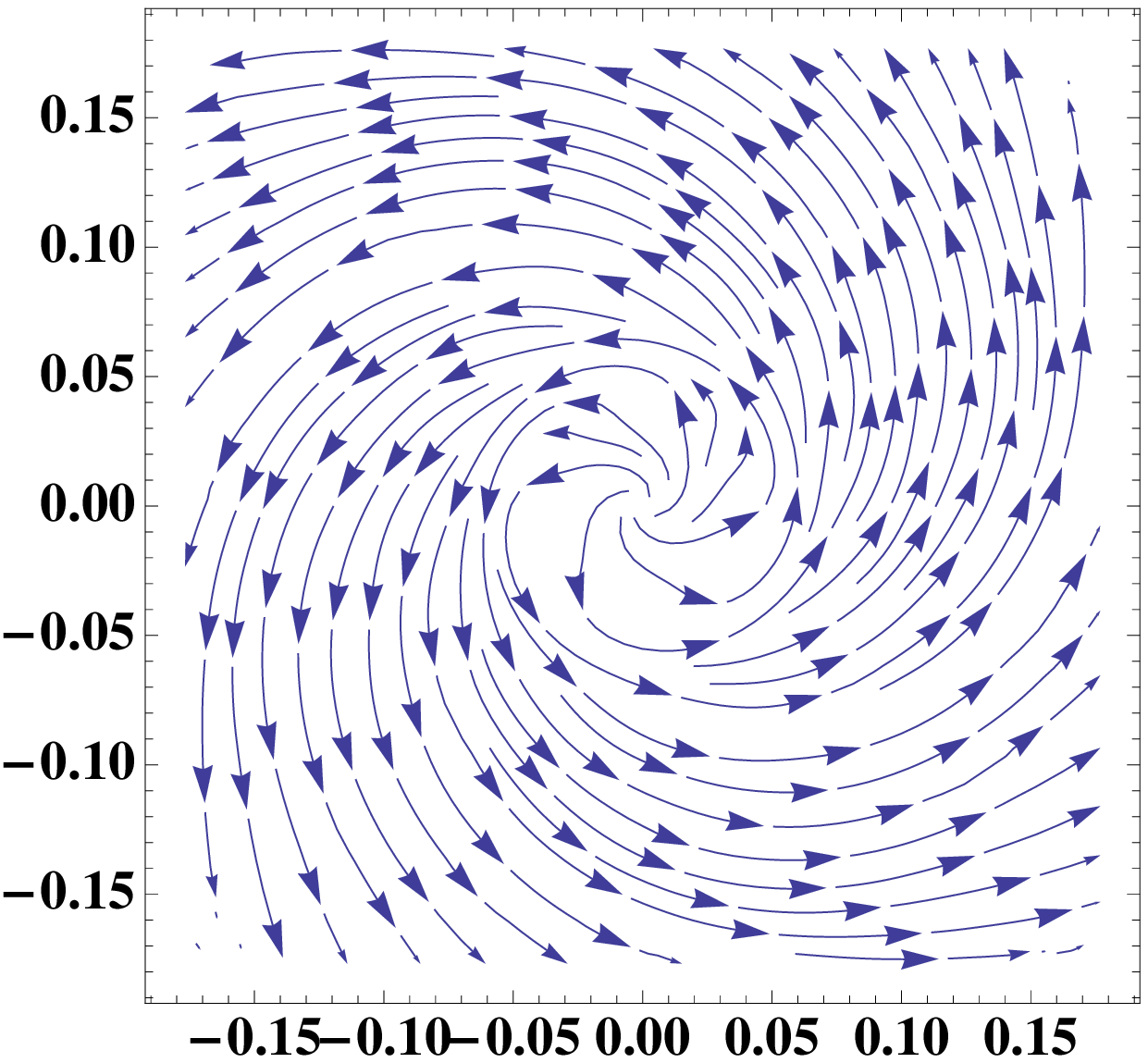}\\
\includegraphics[width=0.3\textwidth,angle=0]{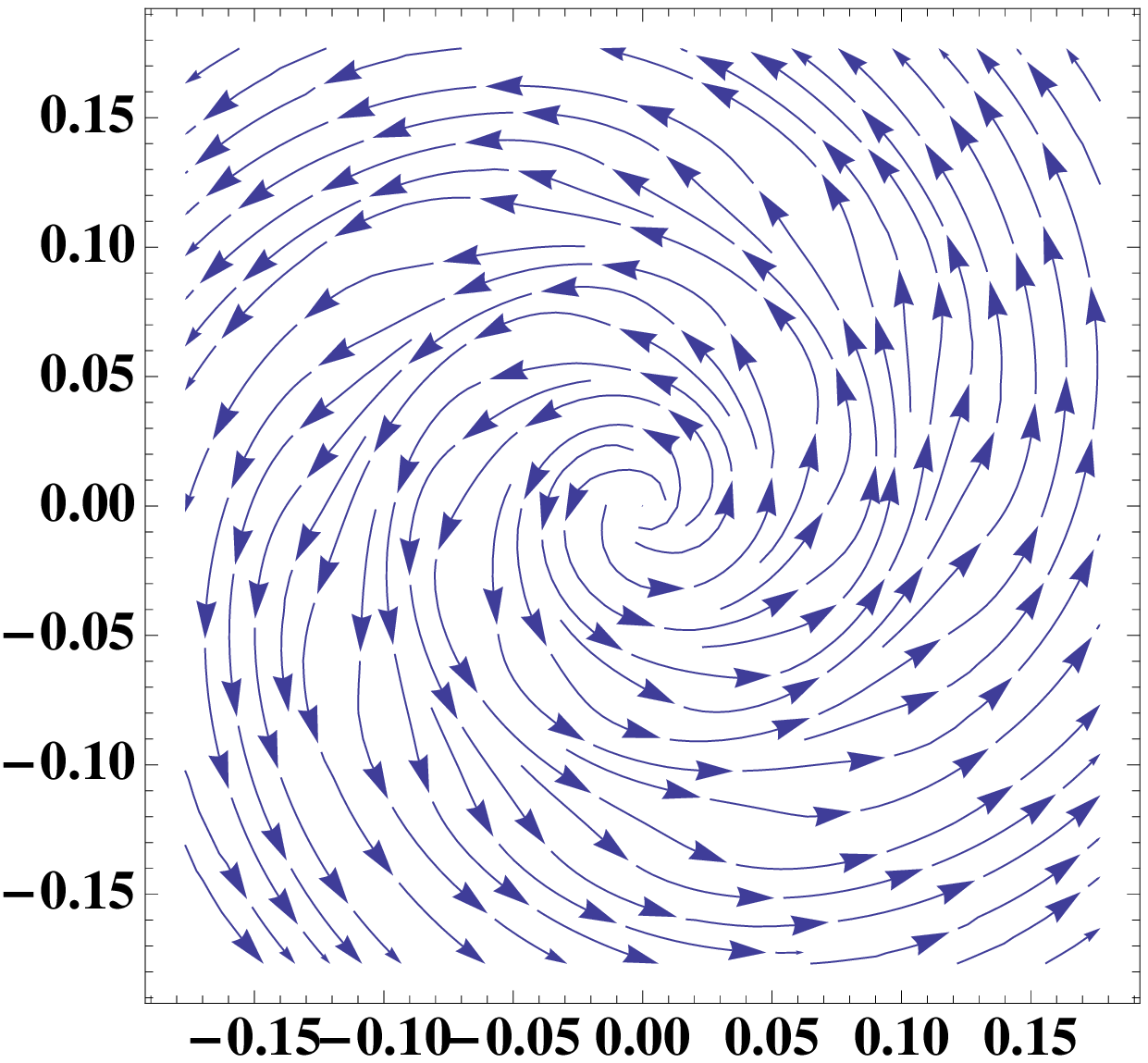}&\hspace{-0.35cm}\includegraphics[width=0.3\textwidth,angle=0]{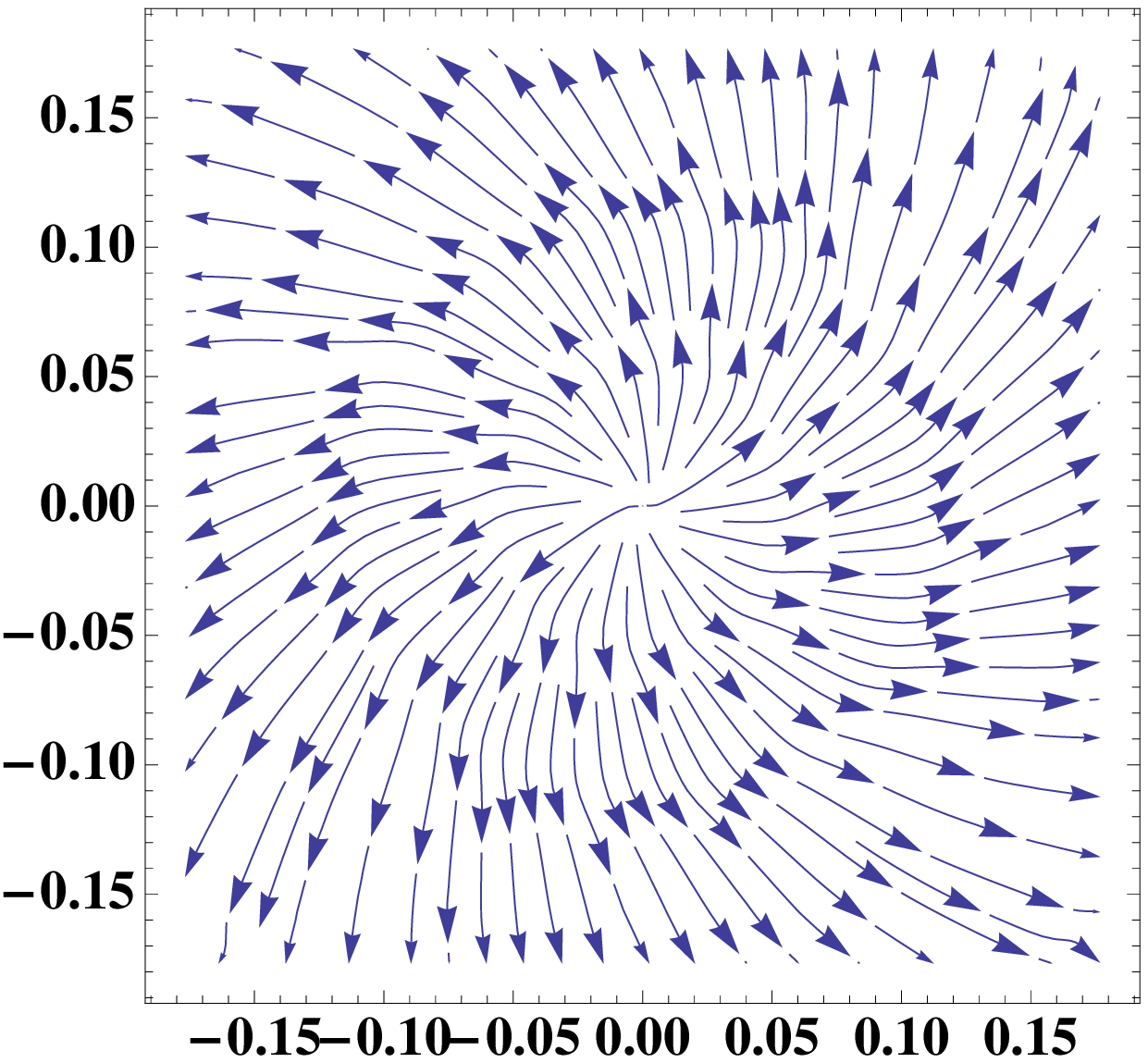}&\hspace{-0.35cm}\includegraphics[width=0.3\textwidth,angle=0]{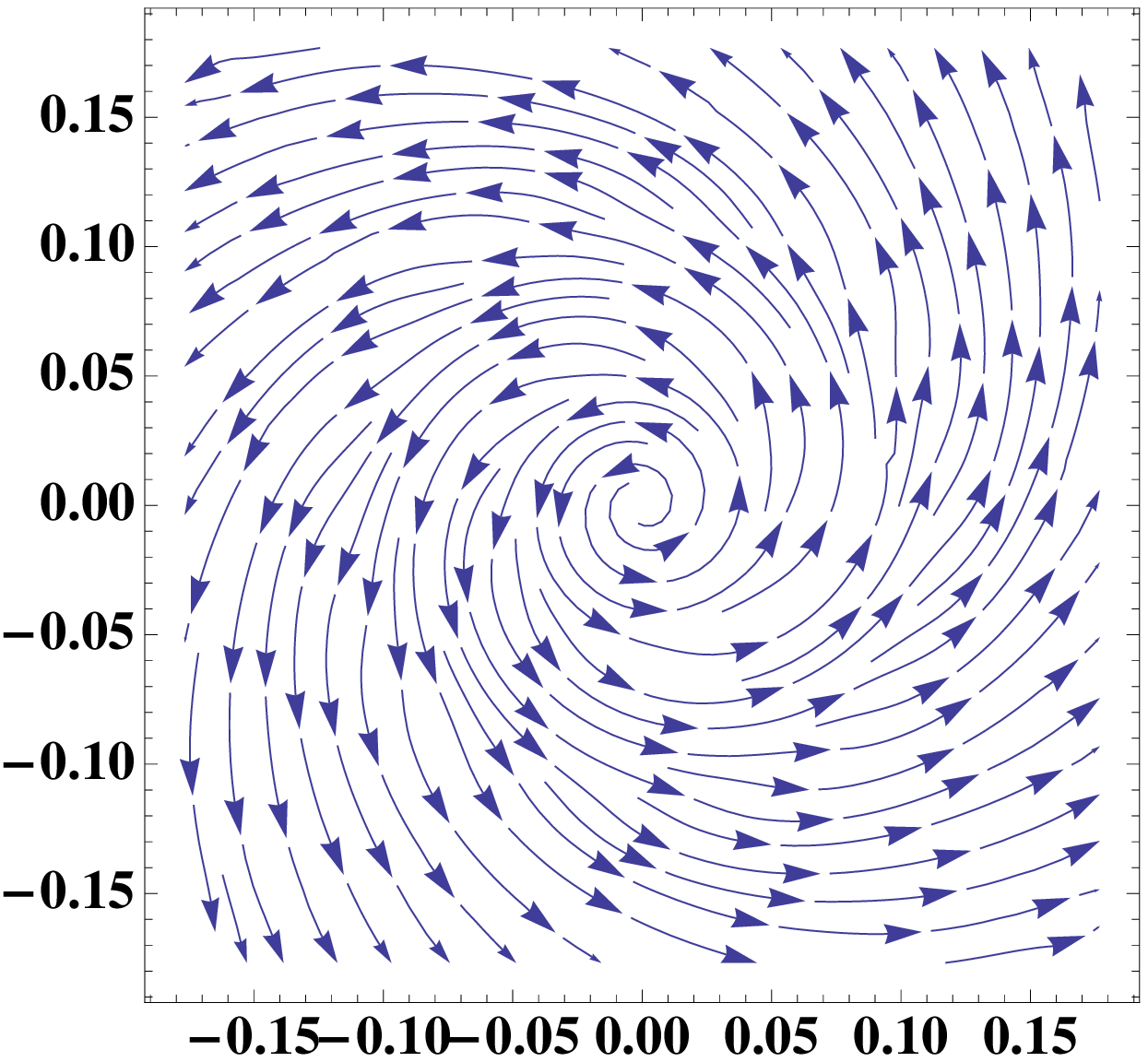}\\
\hspace{0.05cm}\includegraphics[width=0.32\textwidth,angle=0]{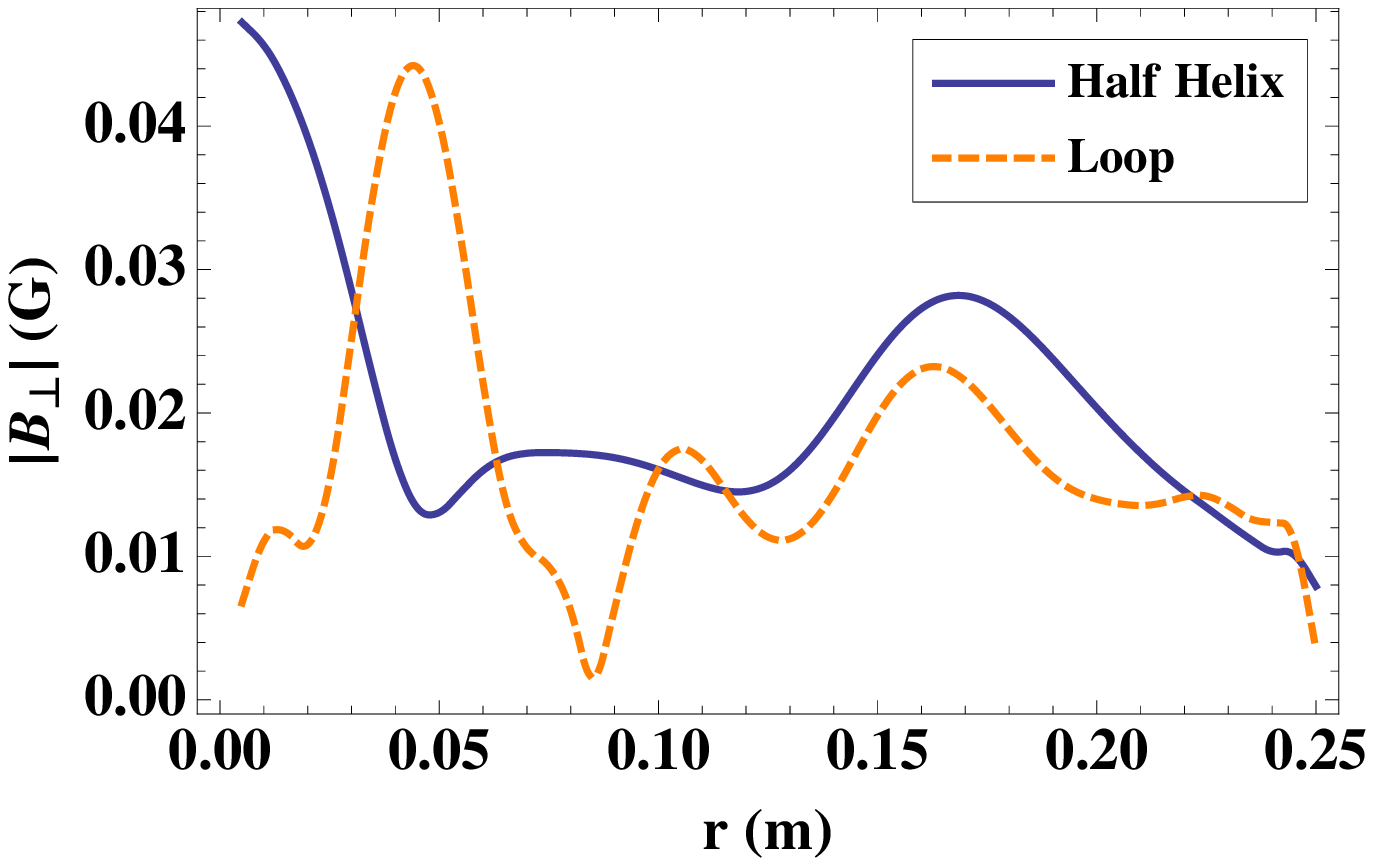}&\includegraphics[width=0.32\textwidth,angle=0]{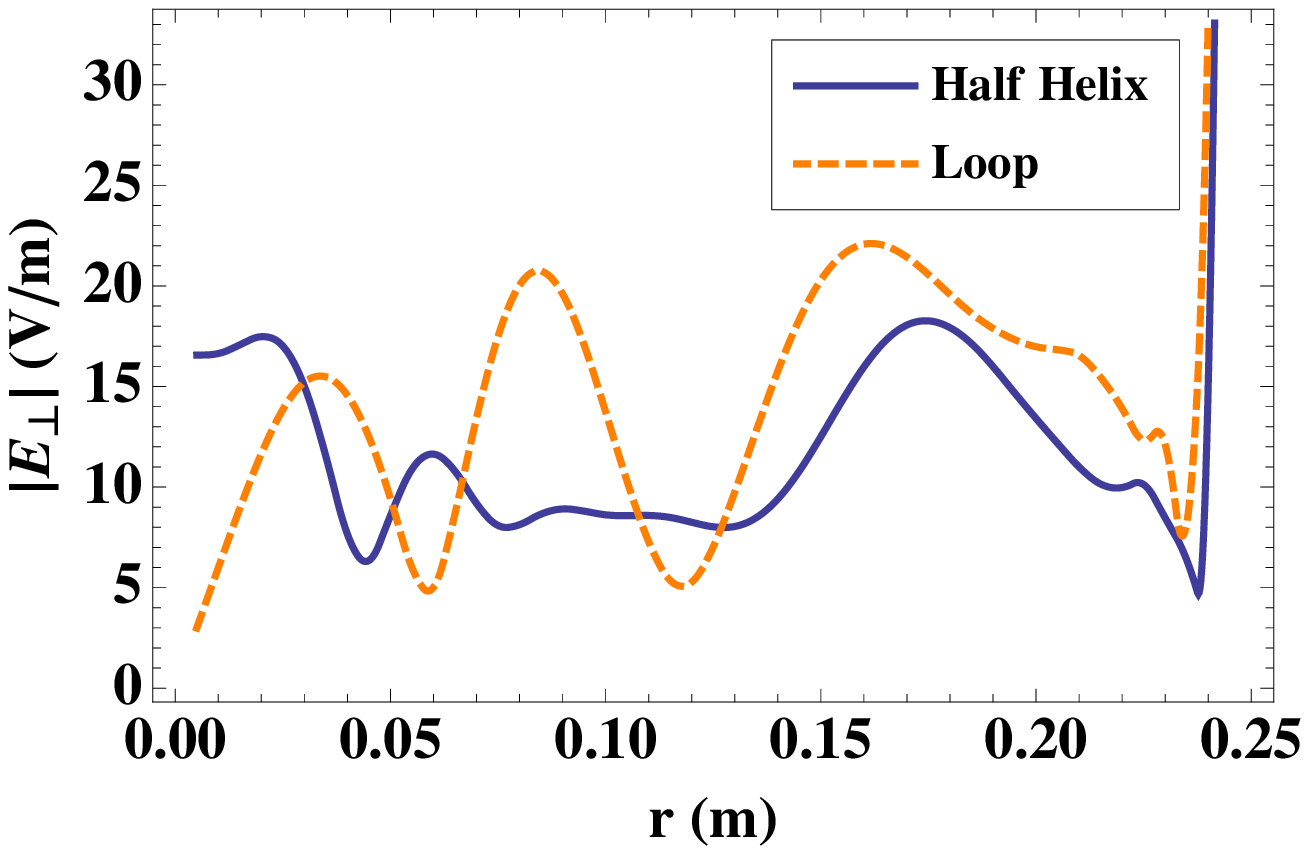}&\hspace{-0.2cm}\includegraphics[width=0.32\textwidth,angle=0]{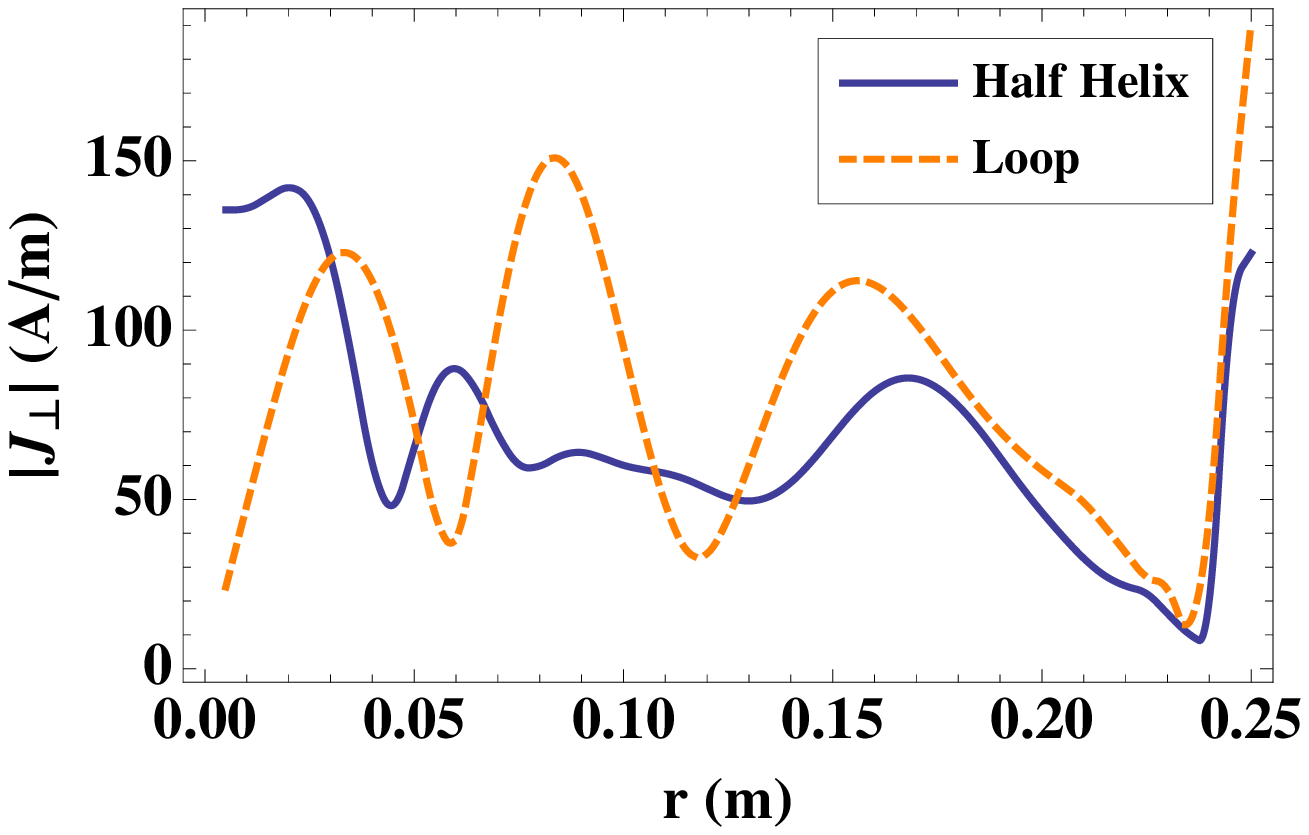}
\end{array}$
\end{center}
\caption{Perpendicular structure ($z=0.2$~m) of wave magnetic field, wave electric field and perturbed current for half helix (first row) and loop (second row) antennas and their comparisons (third row) in terms of magnitude for parabolic density profile.}
\label{fg3}
\end{figure}

We first study the relative power absorption at each radius for half helix, Boswell (or double saddle), Nagoya III and loop antennas. The driving frequency is $13.56$~MHz. Figure~\ref{fg2} displays the computed results for parabolic and Gaussian density profiles shown in Fig.~\ref{fg1}. For radial plots measured at $z=0.2$~m, results have been integrated from $-0.63$~m to $0.63$~m, while for axial plots at $r=0.02$~m, integration has been taken over the plasma cross section. It should be noted that the units of power absorption in radial and axial plots are different, due to different integration dimensions. We can see from the radial plots that the relative power absorption is much bigger near plasma core and edge than that between them, indicating the existence of edge heating and well coupled helicon waves. The off-axis peak of loop antenna features the $m=0$ mode structure which is different from the $m=1$ mode driven by the other three antennas. From the axial plots, we can see that the relative power absorption is symmetric around the driving antenna, except that the half helix antenna exhibits a preferred axial direction. This preference has been observed in various helicon devices and is related to the twisted legs and the direction of static magnetic field\cite{Chen:1996aa, Chen:1996ab, Sudit:1996aa, Lee:2011aa, Chang:2012aa}. Overall, the relative power absorption for parabolic density profile is much bigger than that for Gaussian density profile, and the loop antenna results in higher relative power absorption than the other three antennas. This superiority of loop antenna was also observed before\cite{Soltani:2016aa}.
Figure~\ref{fg3} shows the stream plots of perpendicular wave magnetic field, wave electric field and perturbed current for half helix antenna and loop antenna, together with their comparisons in terms of magnitude. We can see that their perpendicular structures are indeed different, and the loop antenna gives higher magnitudes of $|E_\perp|$ and $|J_\perp|$ between plasma edge and core. The perpendicular structures of Boswell and Nagoya III antennas are nearly the same to those of half helix antenna, and the results of Gaussian density profile show similar perpendicular features compared to those of parabolic density profile. 

\section{Various driving frequencies}\label{freq}
\begin{figure}[ht]
\begin{center}$
\begin{array}{ll}
\includegraphics[width=0.46\textwidth,angle=0]{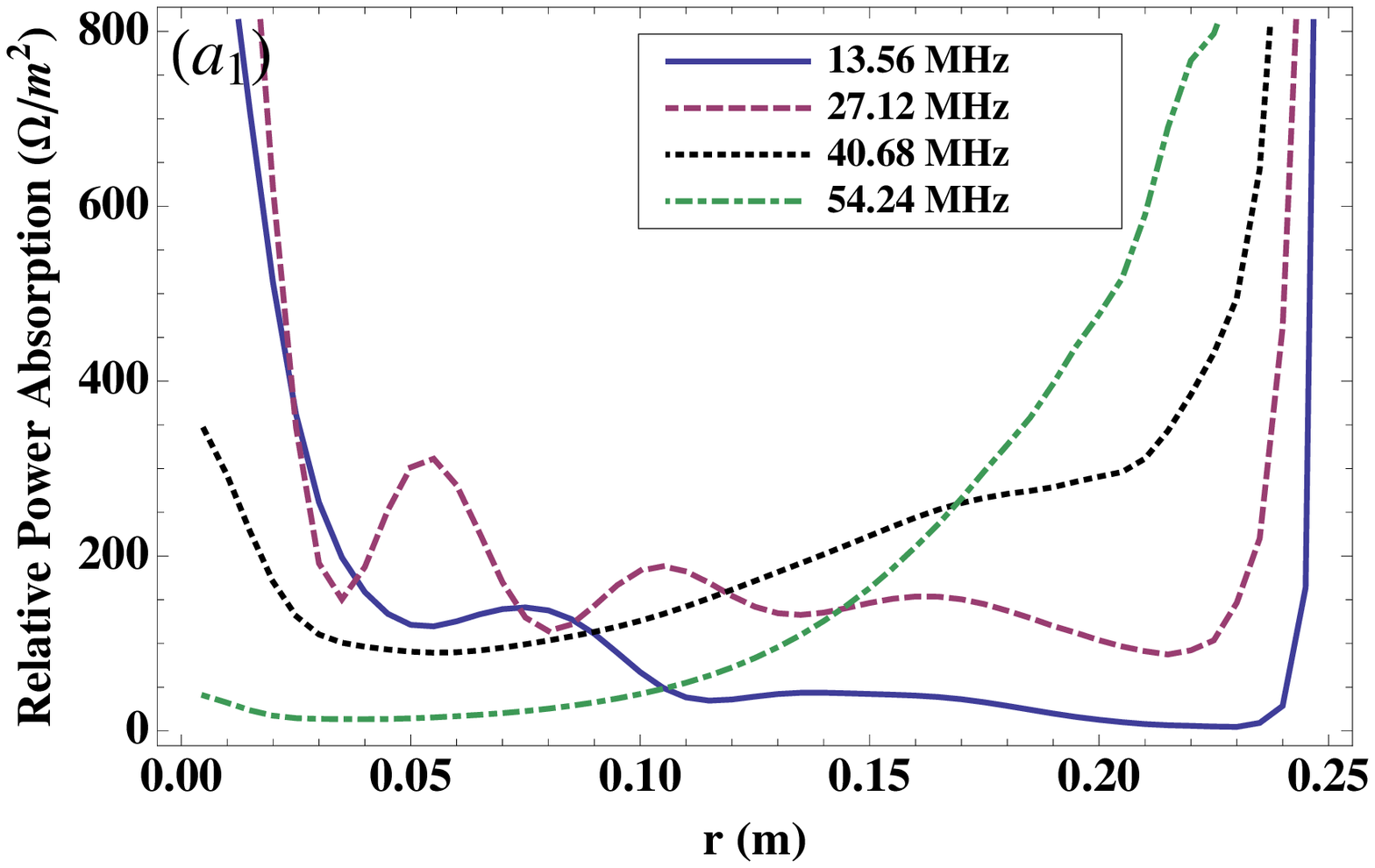}&\hspace{-0.32cm}\includegraphics[width=0.46\textwidth,angle=0]{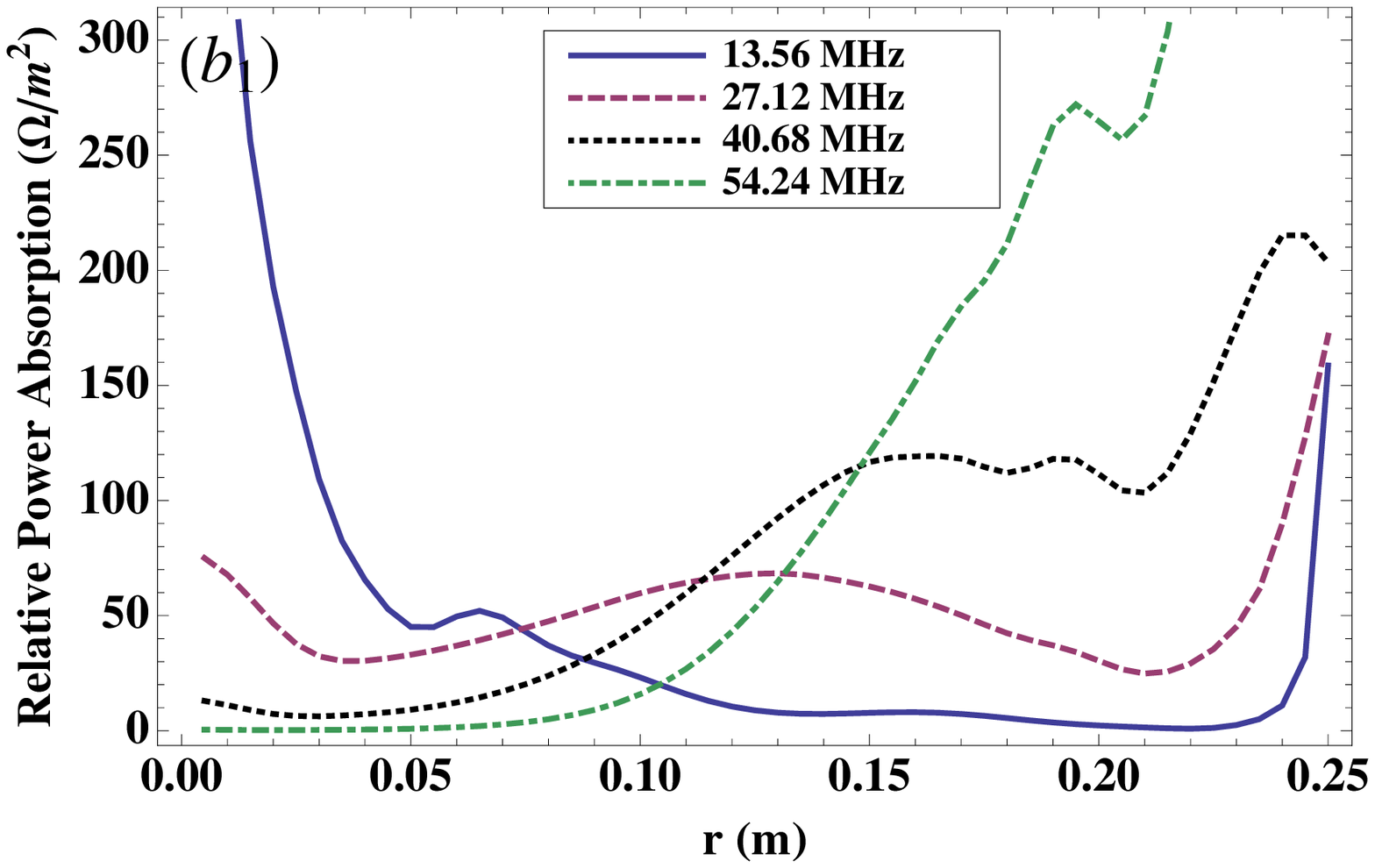}\\
\hspace{0.2cm}\includegraphics[width=0.46\textwidth,angle=0]{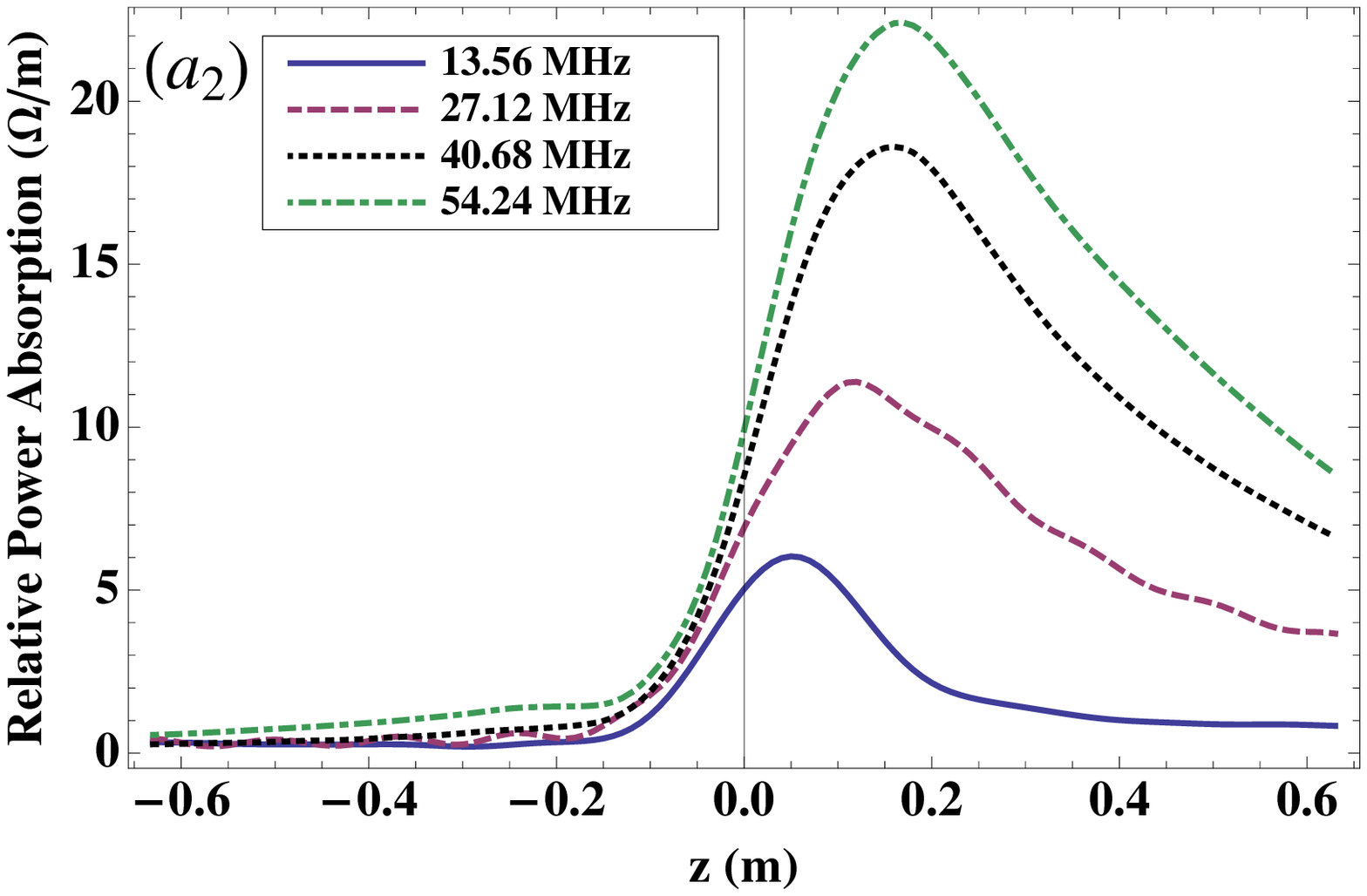}&\includegraphics[width=0.46\textwidth,angle=0]{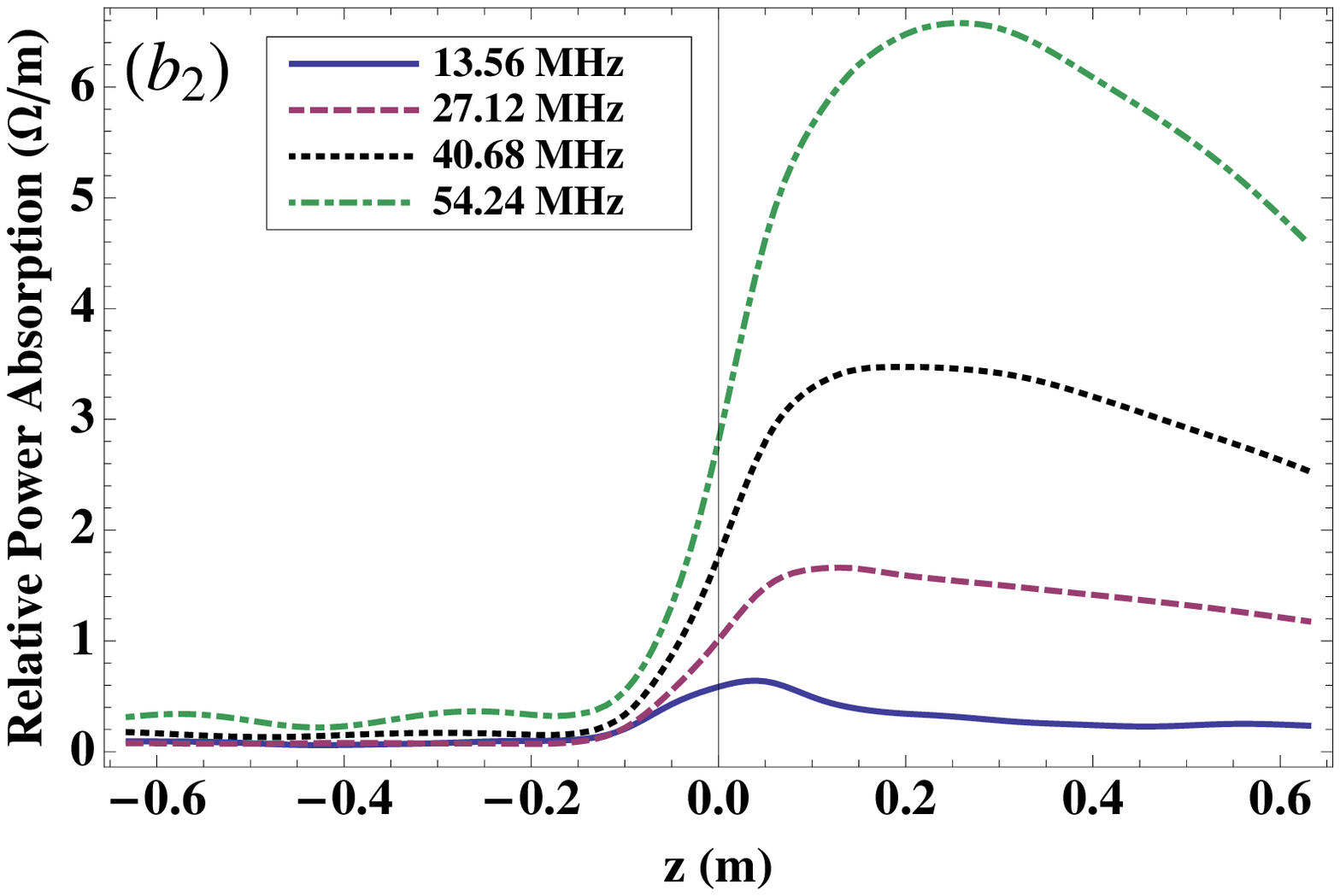}
\end{array}$
\end{center}
\caption{Relative power absorption in radial ($z=0.2$~m) and axial ($r=0.02$~m) directions for various driving frequencies: (a) parabolic density profile, (b) Gaussian density profile.}
\label{fg4}
\end{figure}

\begin{figure}[ht]
\begin{center}$
\begin{array}{ll}
\includegraphics[width=0.46\textwidth,angle=0]{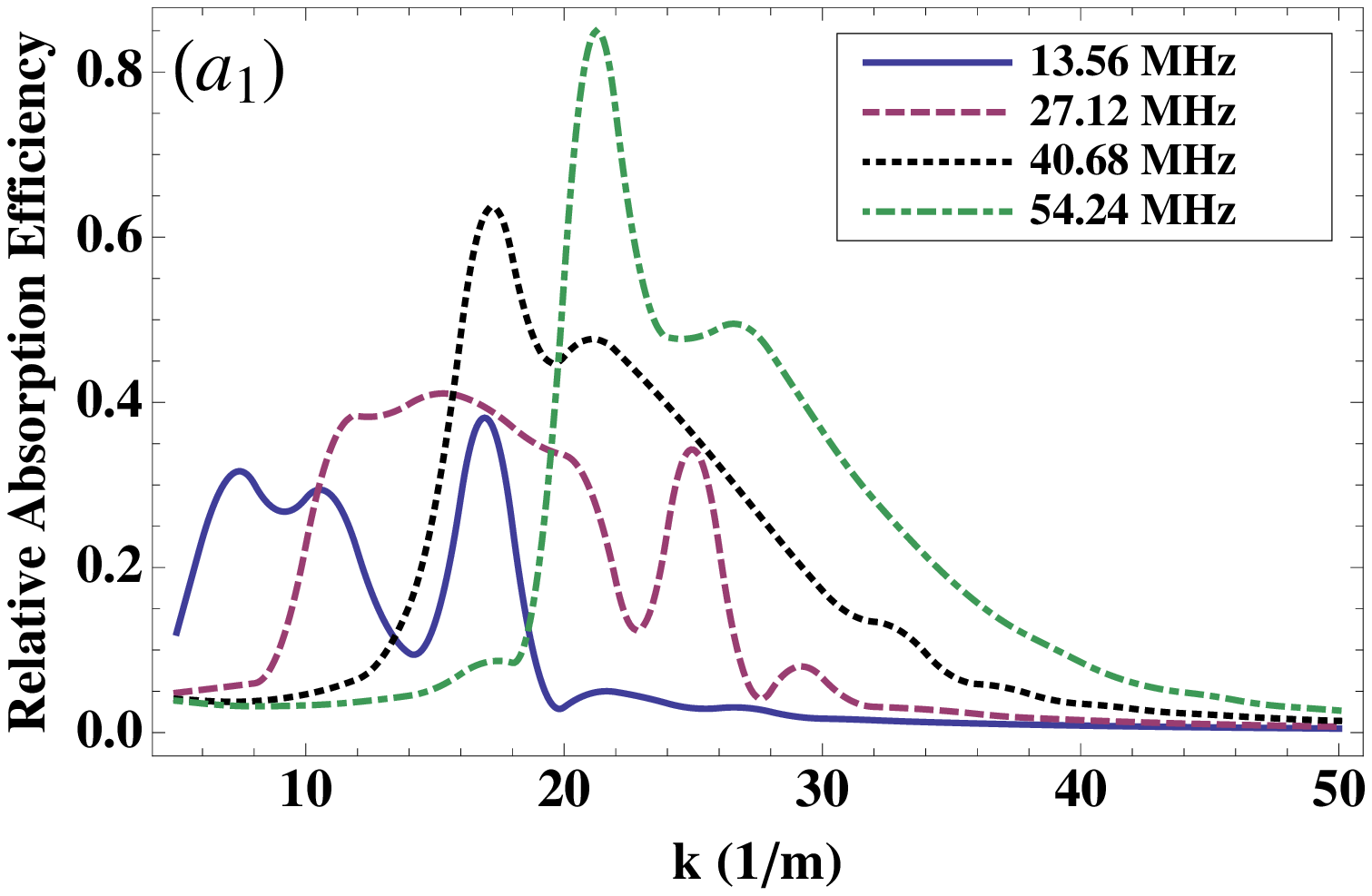}&\includegraphics[width=0.46\textwidth,angle=0]{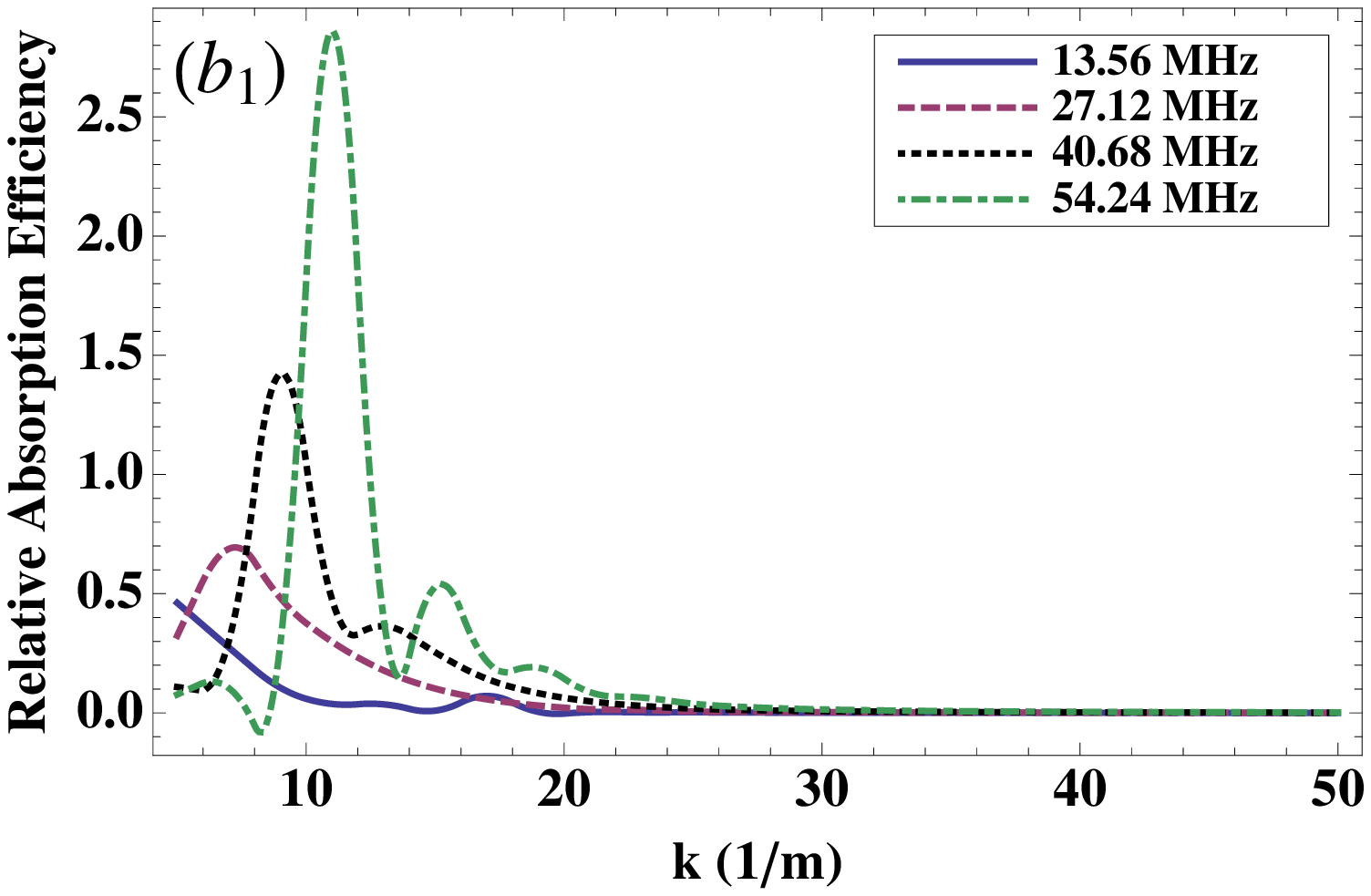}\\
\hspace{0.03cm}\includegraphics[width=0.458\textwidth,angle=0]{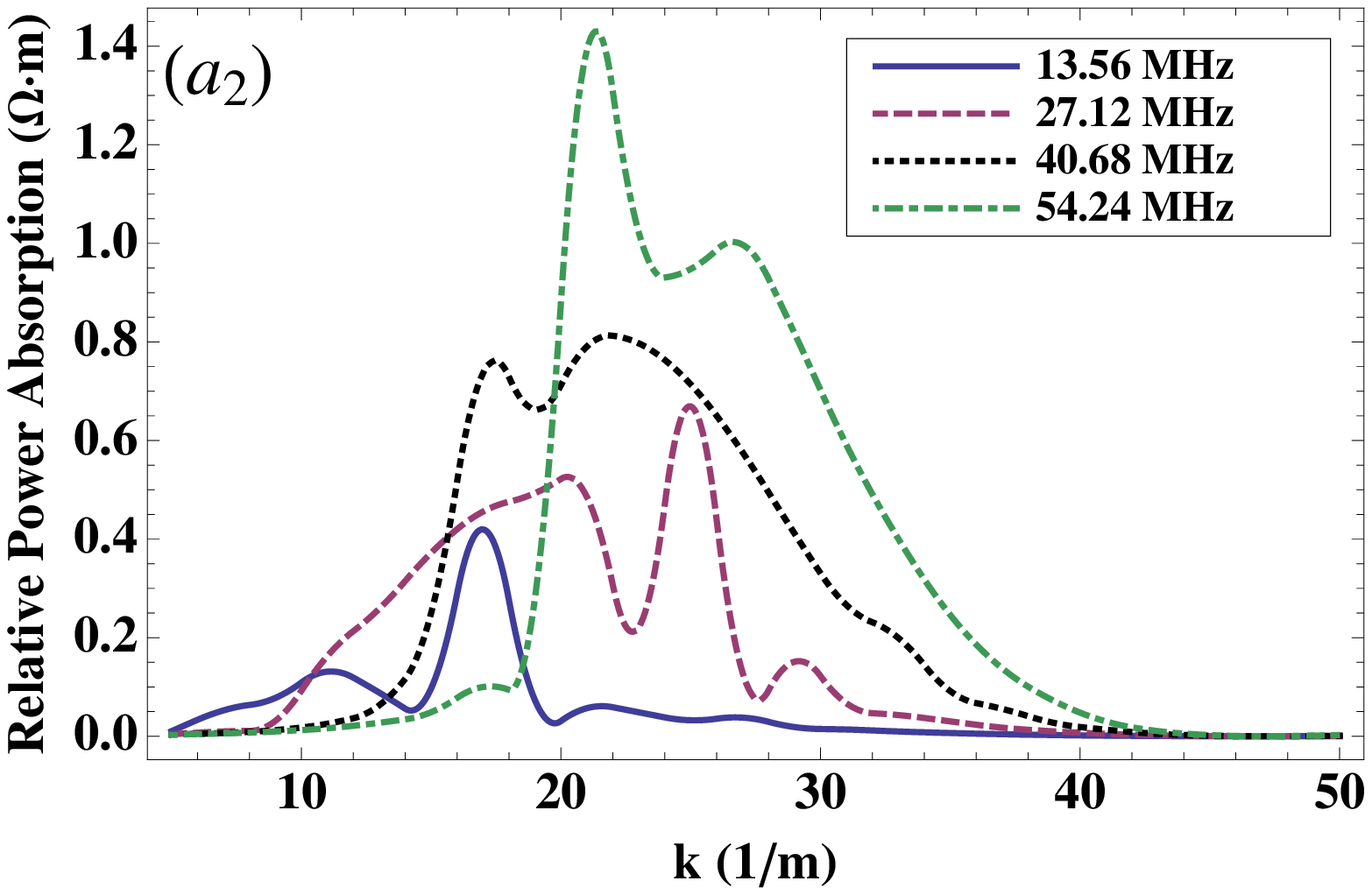}&\includegraphics[width=0.458\textwidth,angle=0]{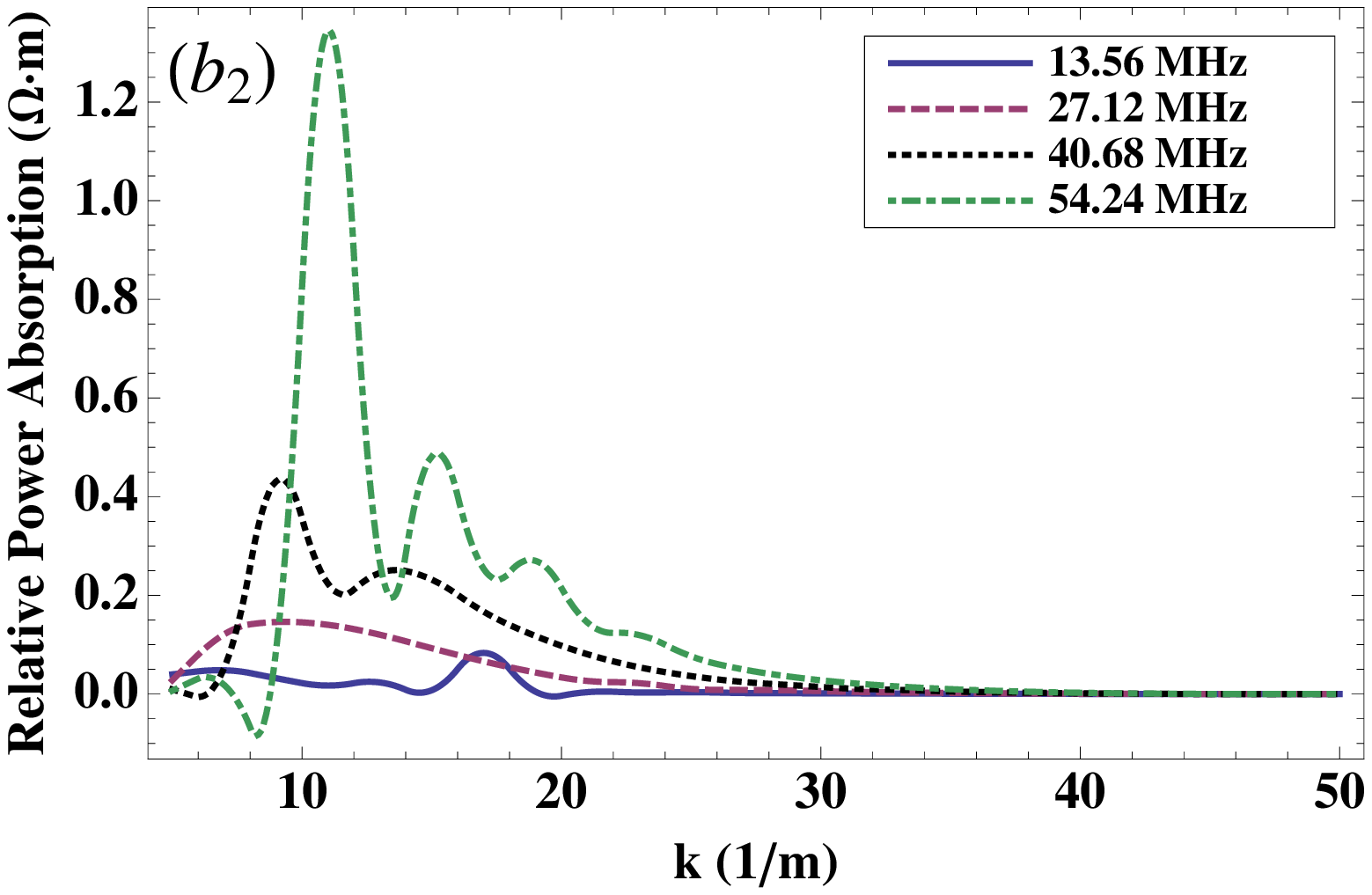}
\end{array}$
\end{center}
\caption{Relative absorption efficiency and relative power absorption as functions of axial wave number: (a) parabolic density profile, (b) Gaussian density profile.}
\label{fg5}
\end{figure}

\begin{figure}[ht]
\begin{center}$
\begin{array}{ccc}
\textbf{27.12 MHz}&\textbf{40.68 MHz}&\textbf{54.24 MHz}\\
\includegraphics[width=0.3\textwidth,angle=0]{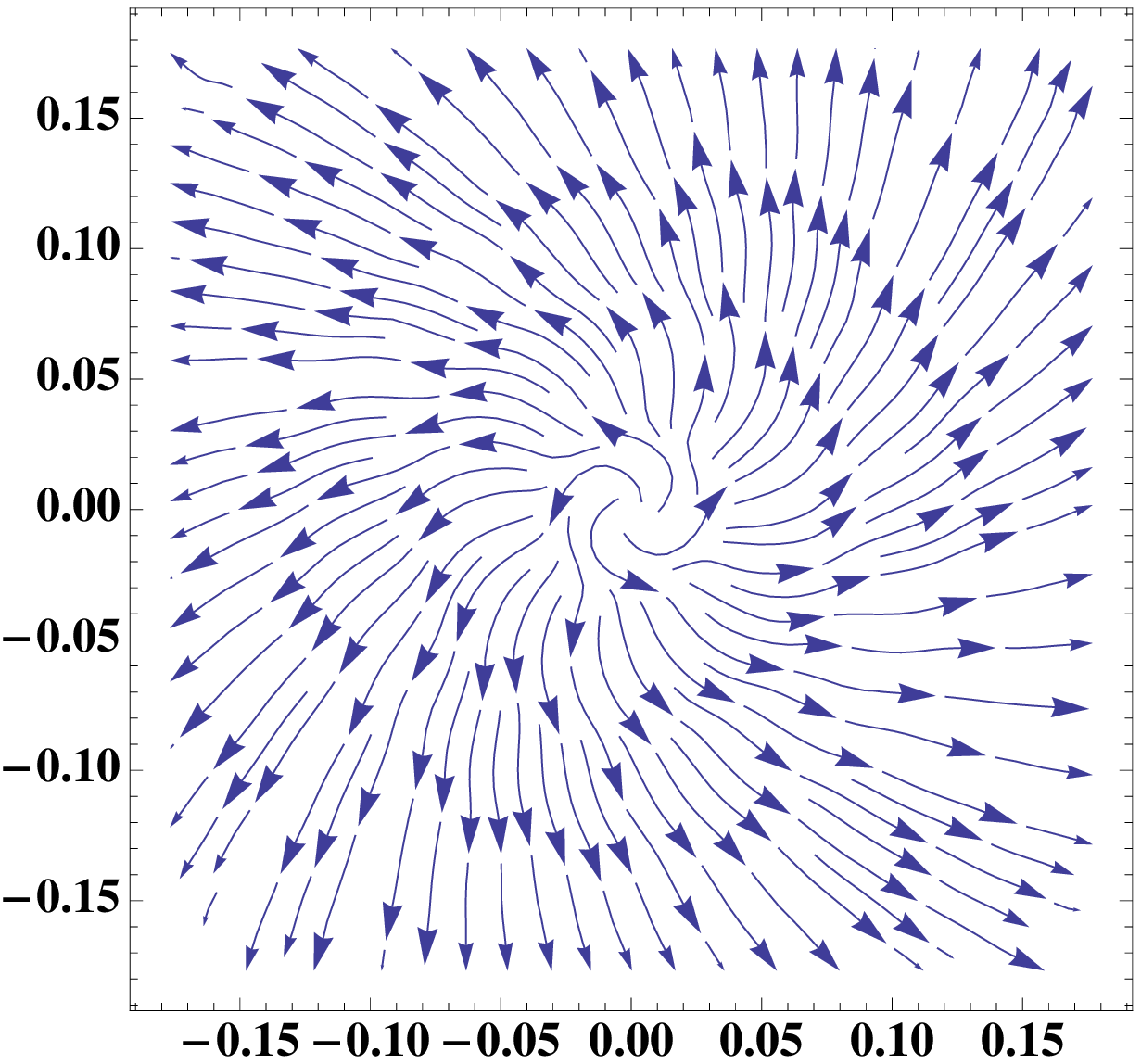}&\includegraphics[width=0.3\textwidth,angle=0]{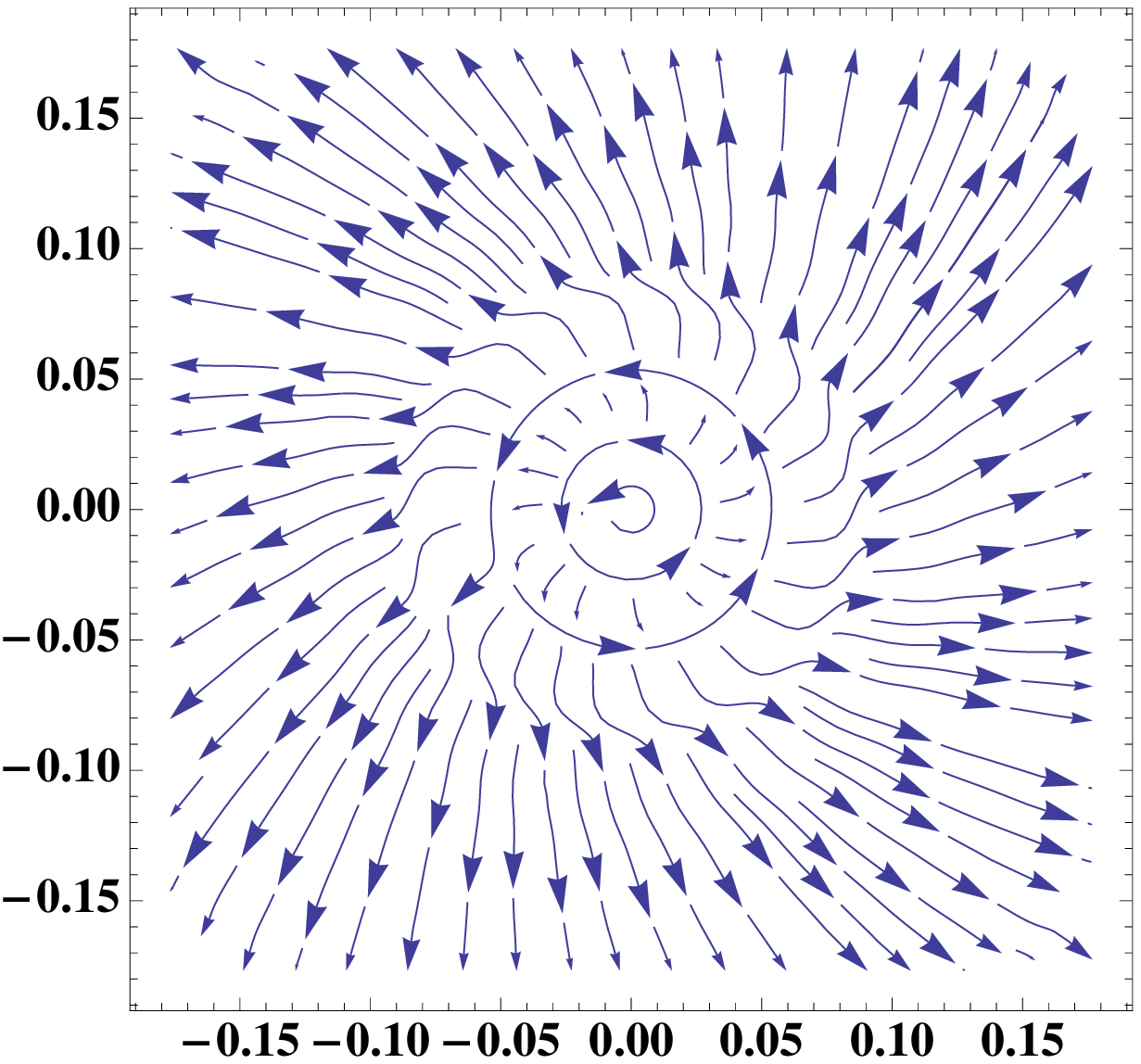}&\includegraphics[width=0.3\textwidth,angle=0]{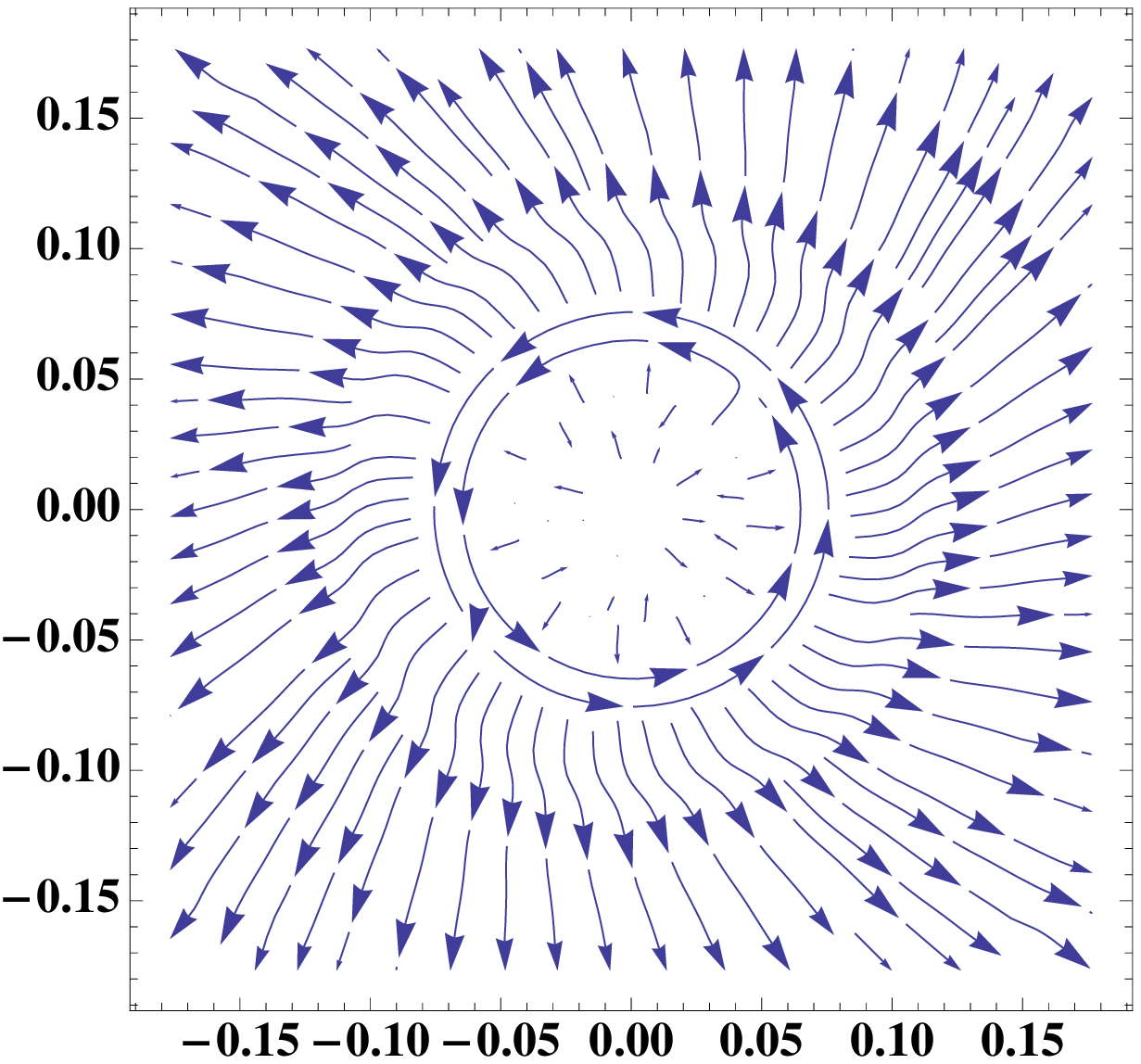}\\
\includegraphics[width=0.3\textwidth,angle=0]{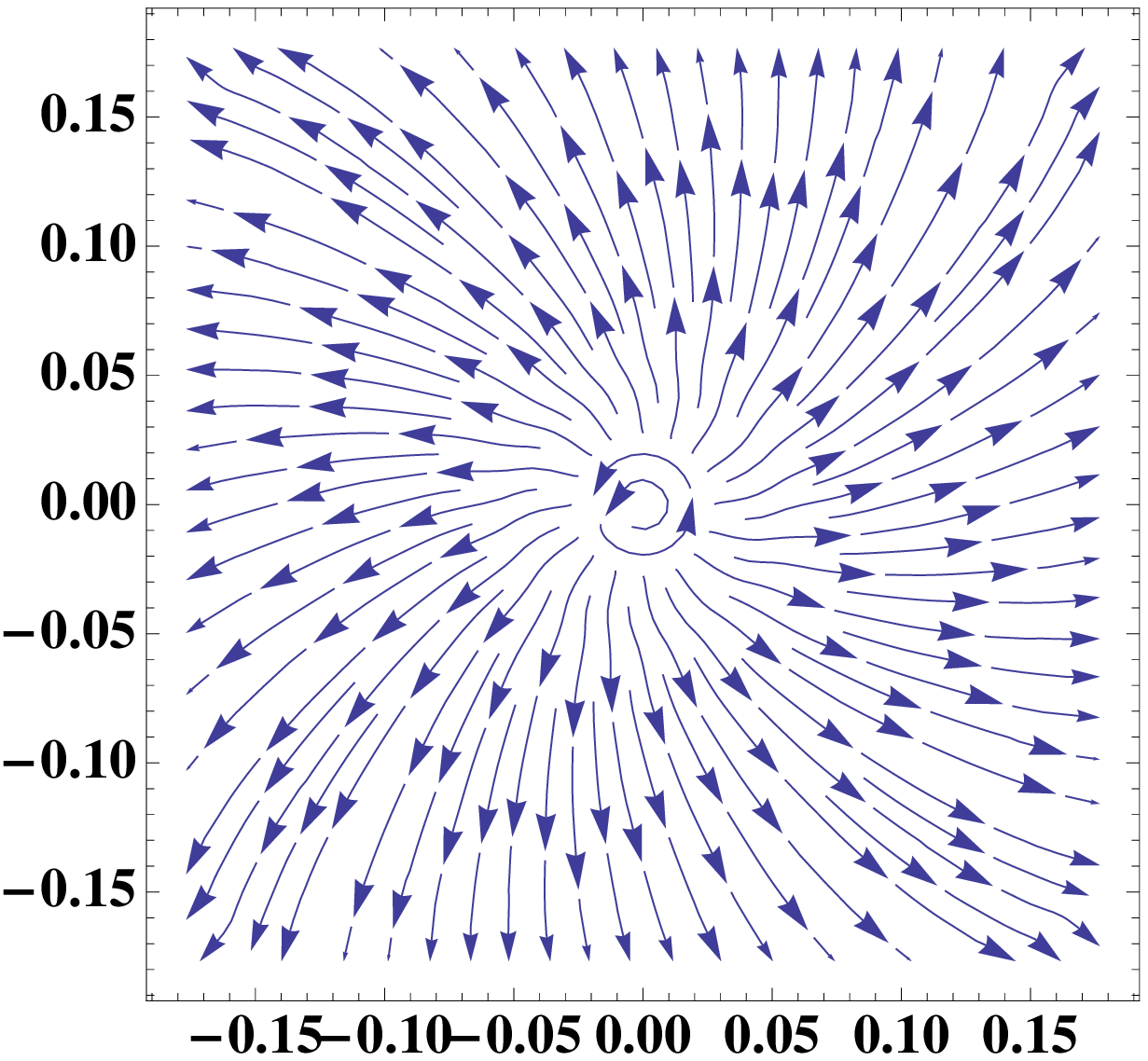}&\includegraphics[width=0.3\textwidth,angle=0]{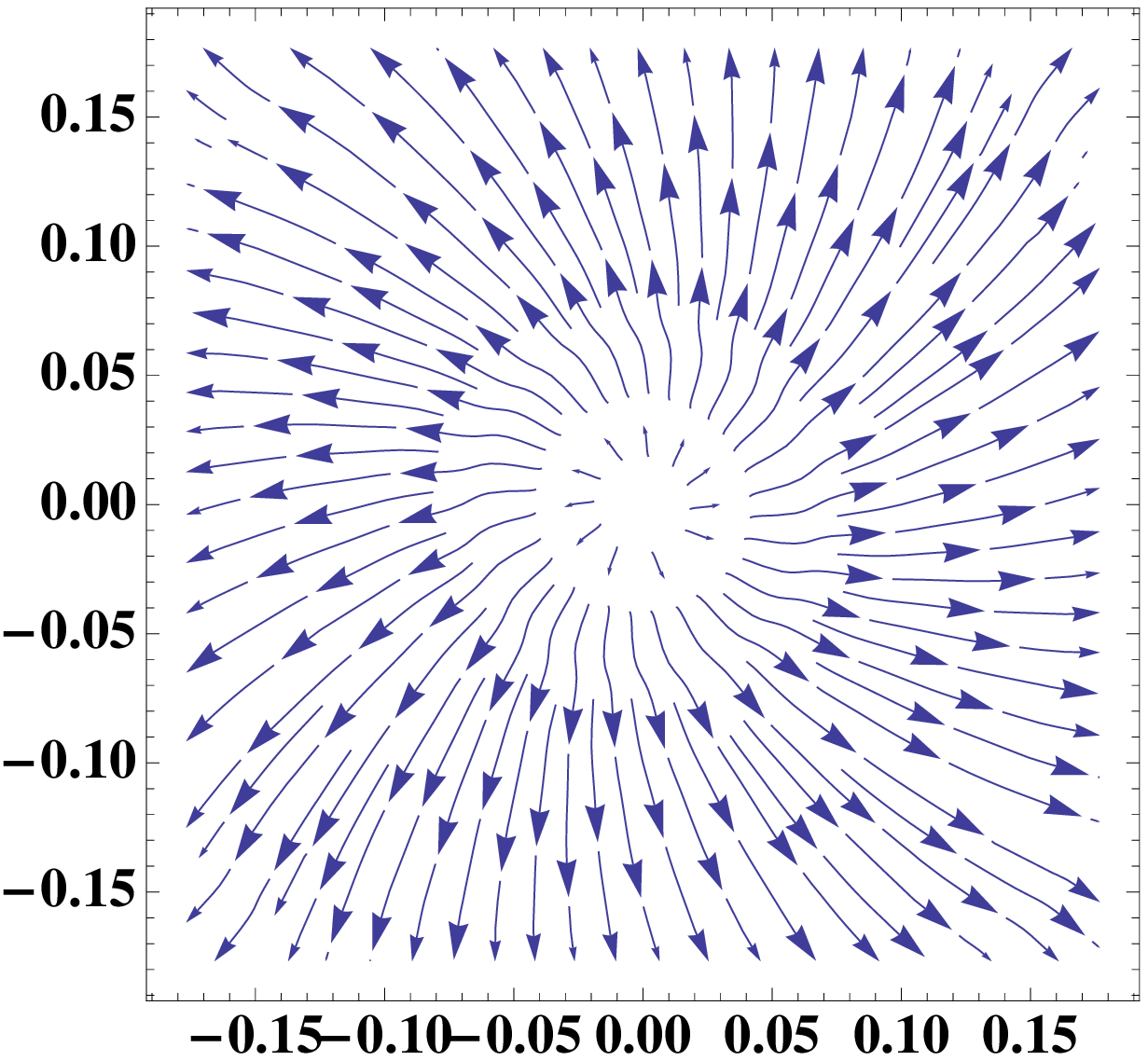}&\includegraphics[width=0.3\textwidth,angle=0]{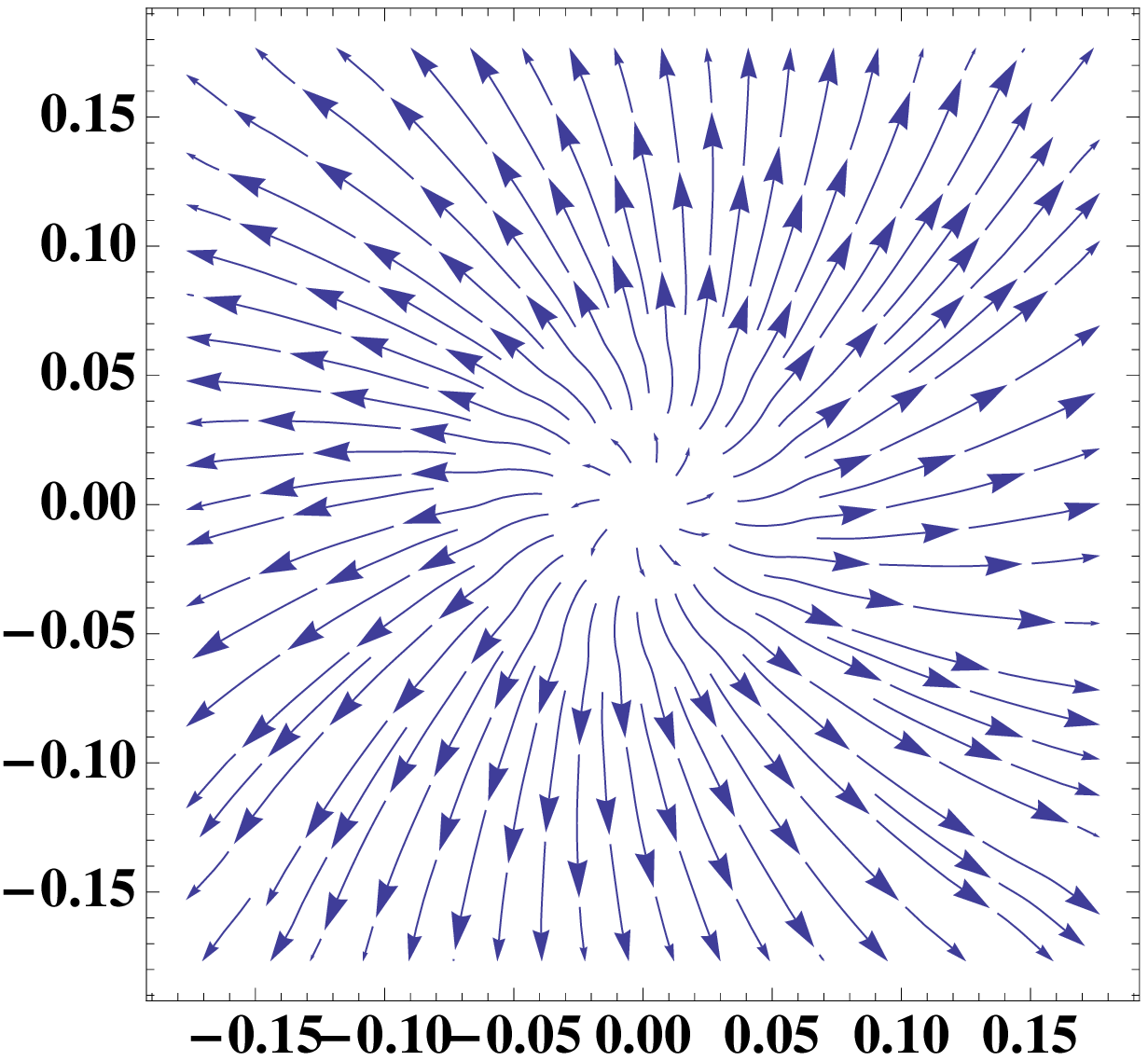}
\end{array}$
\end{center}
\caption{Perpendicular structures of wave electric field for various driving frequencies: upper and lower rows are for parabolic and Gaussian density profiles, respectively.}
\label{fg6}
\end{figure}

\begin{figure}[ht]
\begin{center}$
\begin{array}{lll}
\hspace{0.09cm}\includegraphics[width=0.33\textwidth,angle=0]{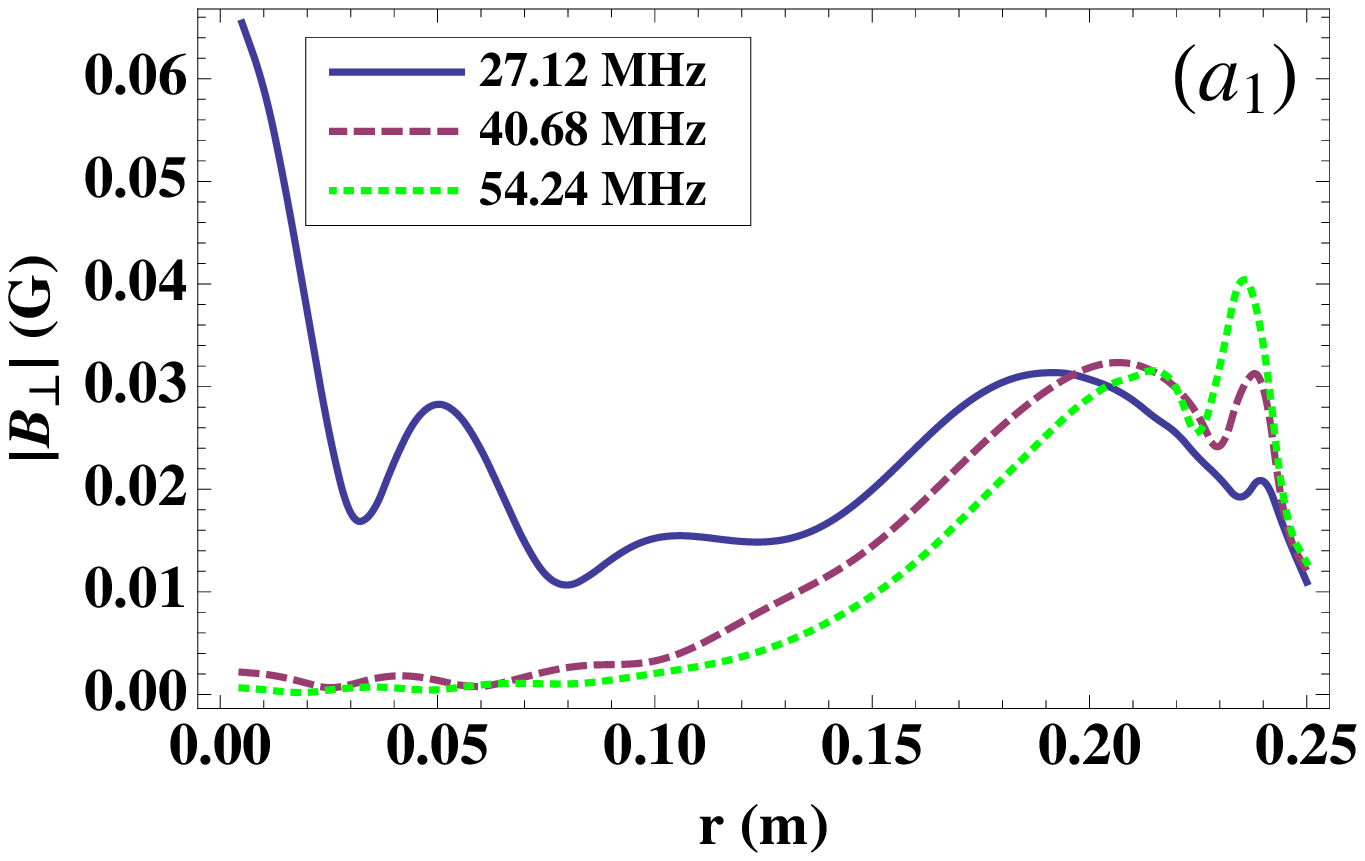}&\hspace{-0.3cm}\includegraphics[width=0.33\textwidth,angle=0]{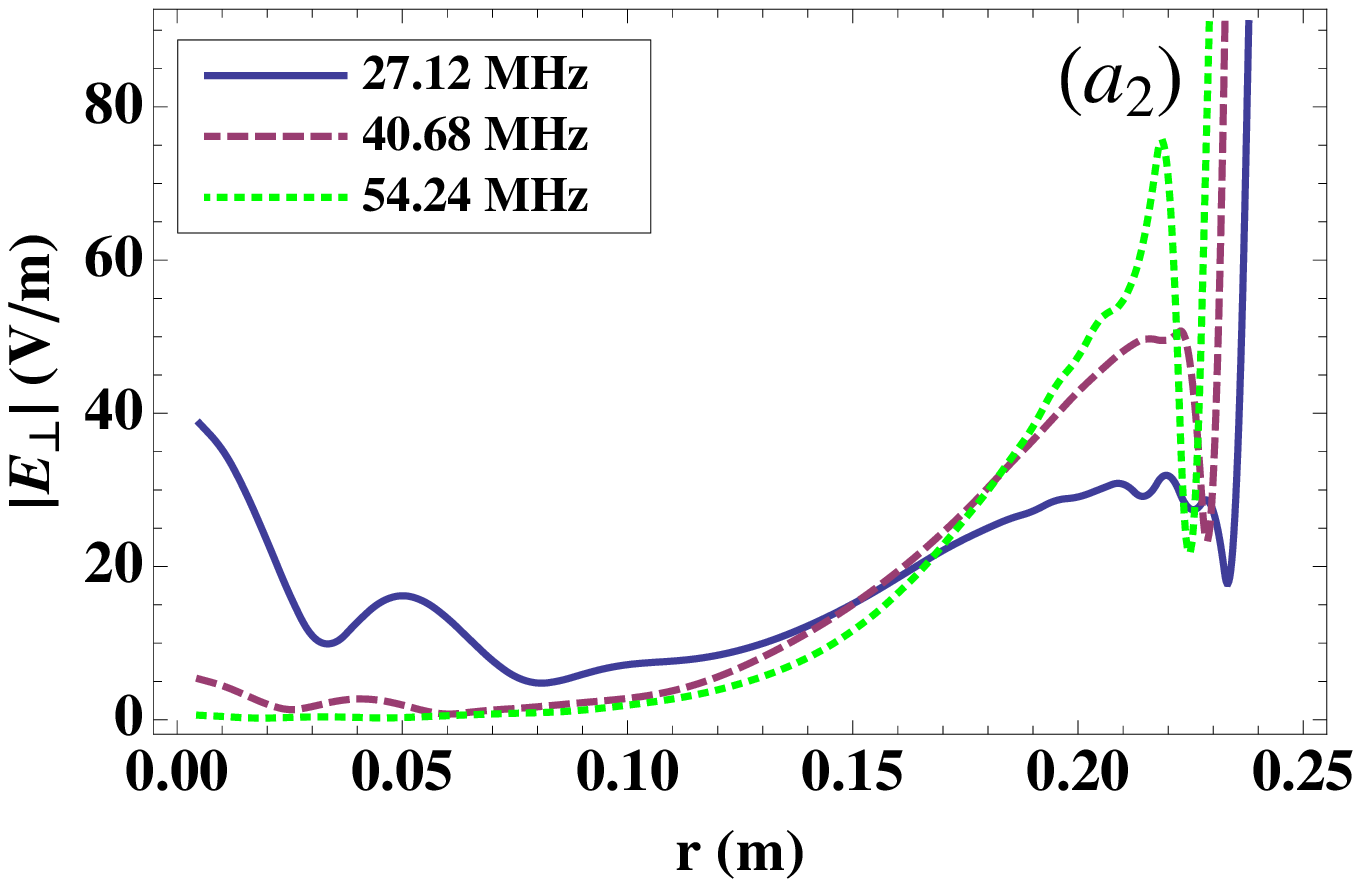}&\hspace{-0.5cm}\includegraphics[width=0.33\textwidth,angle=0]{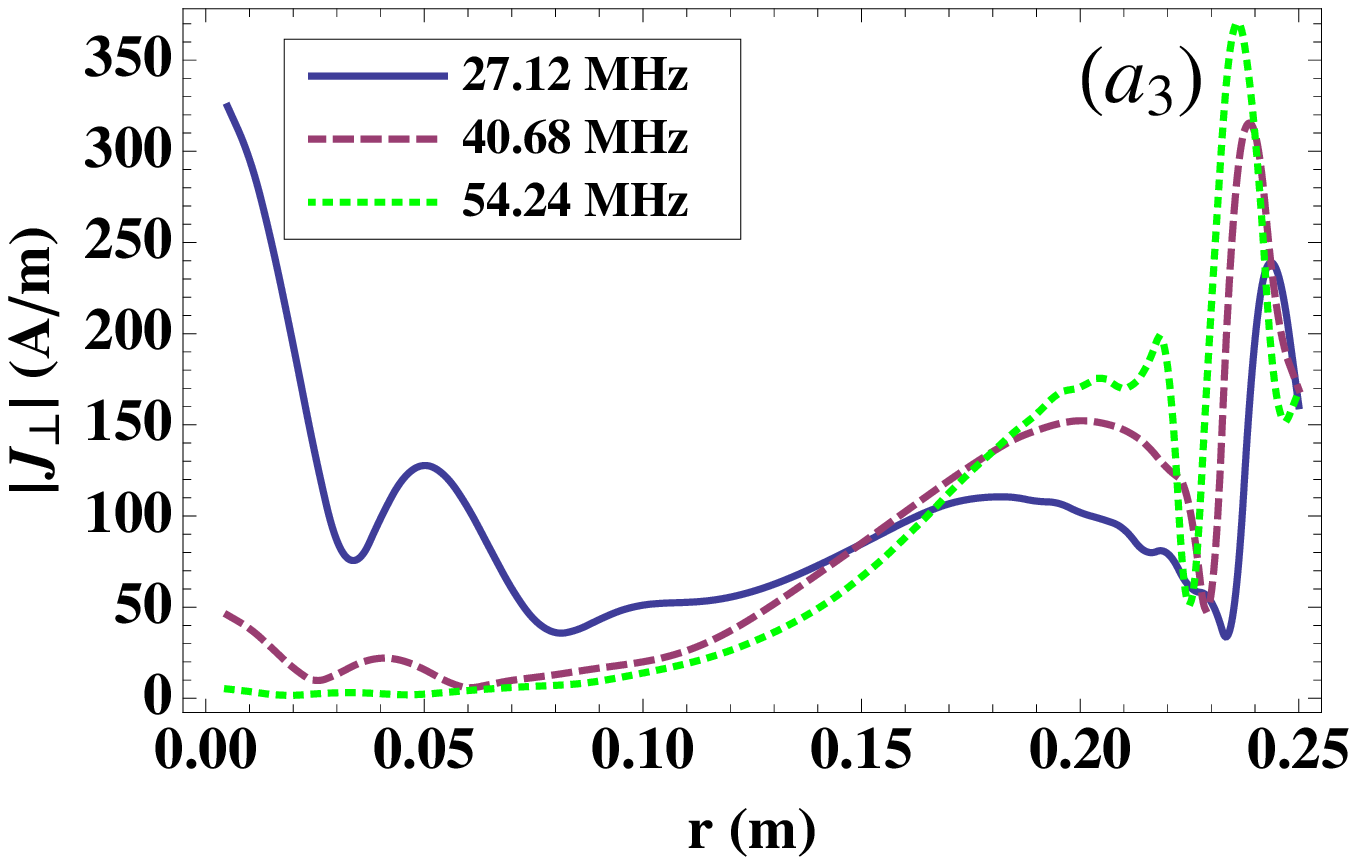}\\
\includegraphics[width=0.33\textwidth,angle=0]{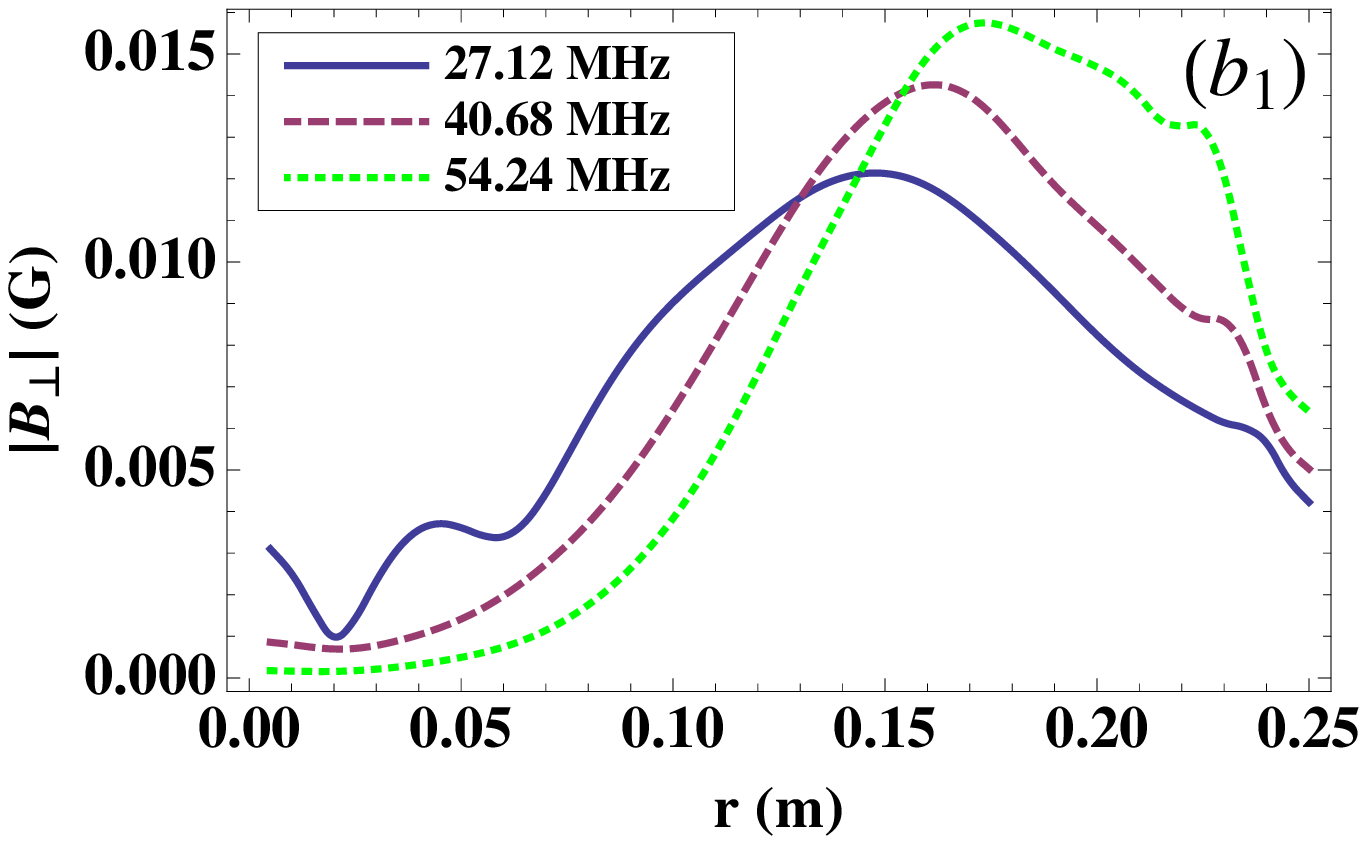}&\hspace{-0.44cm}\includegraphics[width=0.337\textwidth,angle=0]{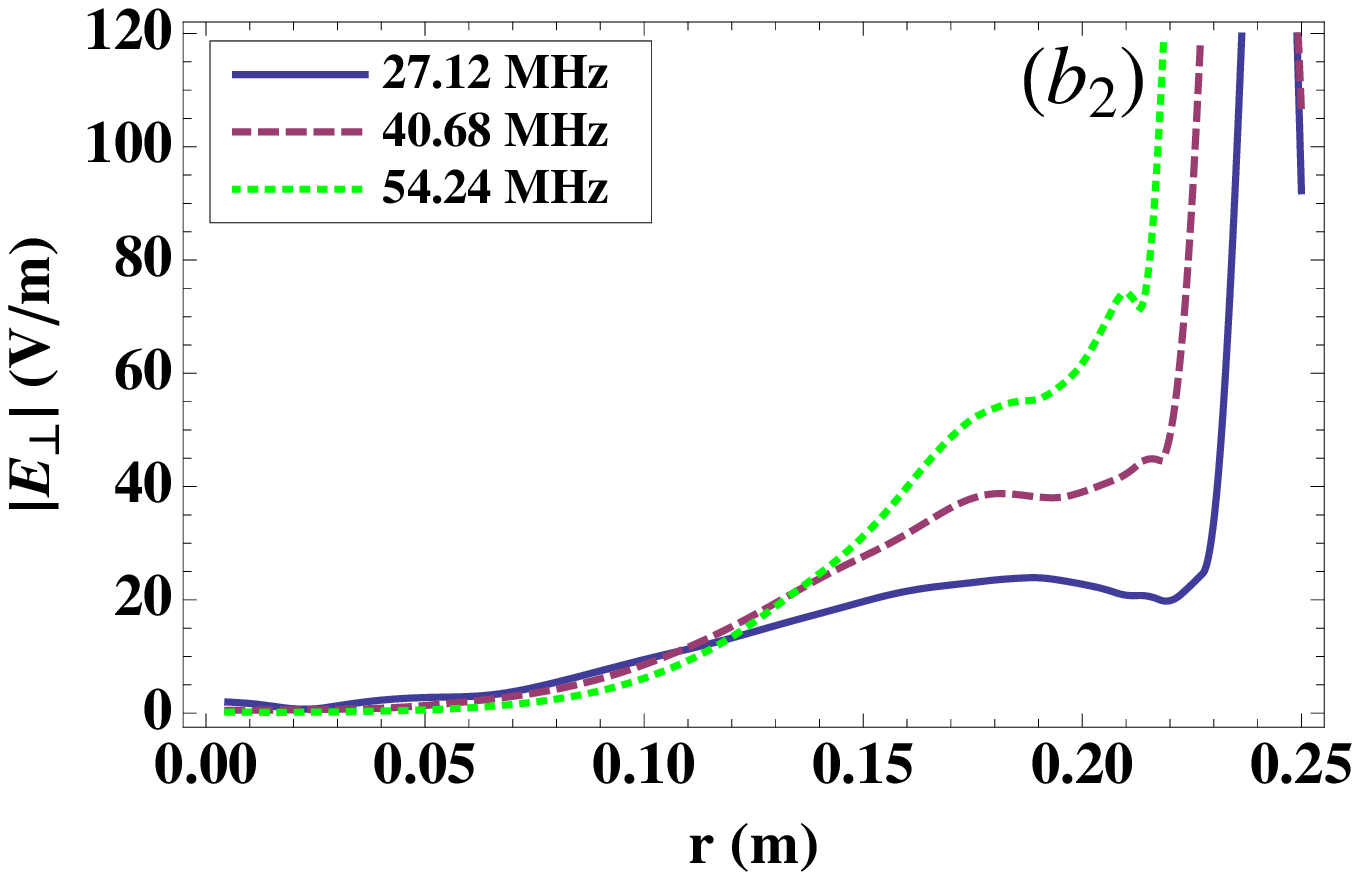}&\hspace{-0.4cm}\includegraphics[width=0.33\textwidth,angle=0]{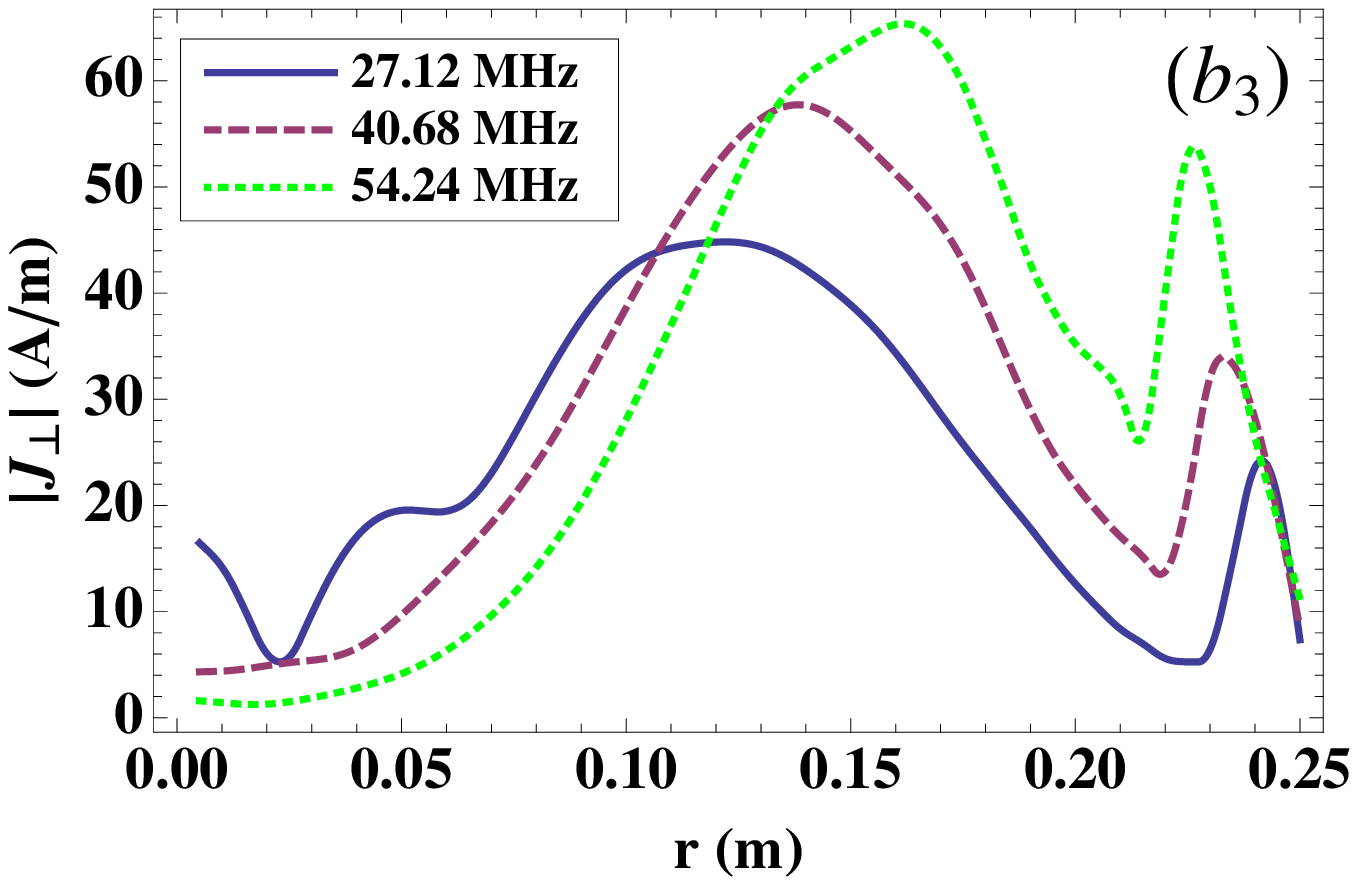}
\end{array}$
\end{center}
\caption{Radial profiles of wave magnetic field, wave electric field and perturbed current for various driving frequencies: (a) parabolic density profile, (b) Gaussian density profile.}
\label{fg7}
\end{figure}

Because driving frequency is a very important parameter for coupling optimization, we choose four typical frequencies and compare their relative power absorptions. Here, we consider the widely used half helix antenna as an example. As shown in Fig.~\ref{fg4}, the relative power absorption increases significantly near plasma edge when the antenna driving frequency becomes higher, and correspondently it drops near plasma core in the same time. This implies that Trivelpiece-Gould mode attracts much more power from antenna than helicon mode at high frequencies, \textcolor{blue}{possibly due to fast electron motion and hence strong electrostatic heating. Decreasing the background pressure and thereby collisional damping to increase electron free path may help redistribute the power absorption from edge towards core}. The axial plots exhibit a more clear trend that the relative power absorption increases with the driving frequency. Francis Chen\cite{Chen:2012aa} obtained the same trend through simulations. Again, the parabolic density profile leads to overall bigger power absorption than the Gaussian density profile. 
This increased power absorption with higher frequency is also confirmed in the spectral space of $k$. Figure~\ref{fg5} gives the computed relative absorption efficiency and relative power absorption in the range of $5<k<50$ for various driving frequencies. Please note that here the absorption efficiency is also normalized to antenna current of $1$~A thereby can be bigger than $100$\%. The peak absorption shifts to bigger value of $k$, as well as the overall absorption grows, for higher driving frequencies. Moreover, for parabolic density profile, efficient power absorption occurs over a wide range of $k$, whereas for Gaussian density profile it is more localized around small values of $k$.
The perpendicular structures of wave electric field for $27.12$~MHz, $40.68$~MHz and $54.24$~MHz are illustrated in Fig.~\ref{fg6}, for both parabolic and Gaussian density profiles. We can see that they change remarkably with increasing frequency, especially for the parabolic density profile. Further, as the frequency increases, the rotation near plasma core becomes much stronger and the rotational area becomes larger, which may be the reason for enhanced power absorption. Their magnitudes are compared in Fig.~\ref{fg7}, together with the magnitudes of wave magnetic field and perturbed current. A similar conclusion can be drawn as from Fig.~\ref{fg4} that more power is absorbed near plasma edge when the driving frequency becomes higher.  Additionally, the perpendicular structures of wave magnetic field and perturbed current do not change noticeably in the range of $13.56-70$~MHz and are hence not shown here. 

\section{Parameter study of plasma response}\label{scan}
\begin{figure}[ht]
\begin{center}$
\begin{array}{l}
\includegraphics[width=0.49\textwidth,angle=0]{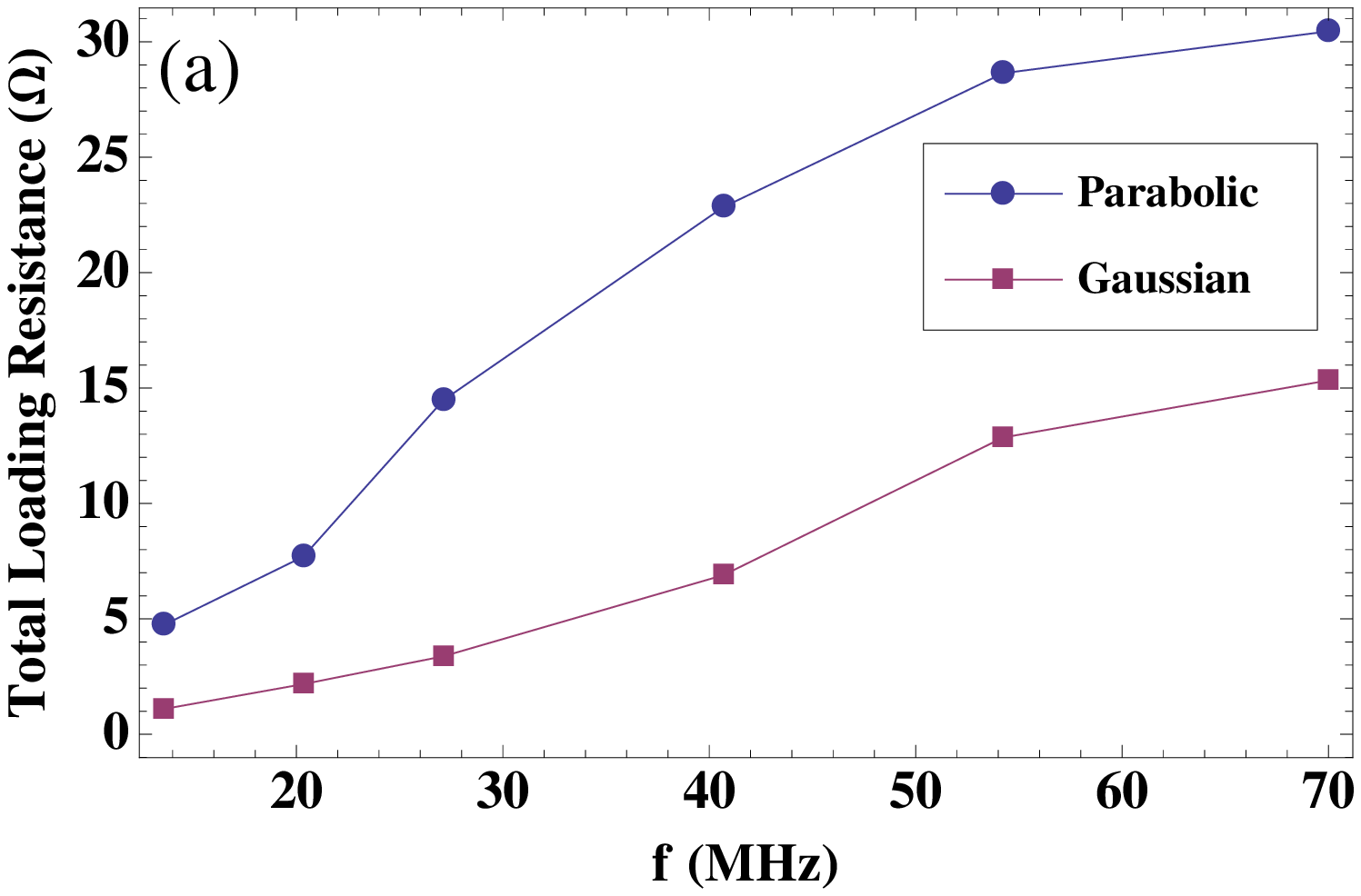}\\
\includegraphics[width=0.49\textwidth,angle=0]{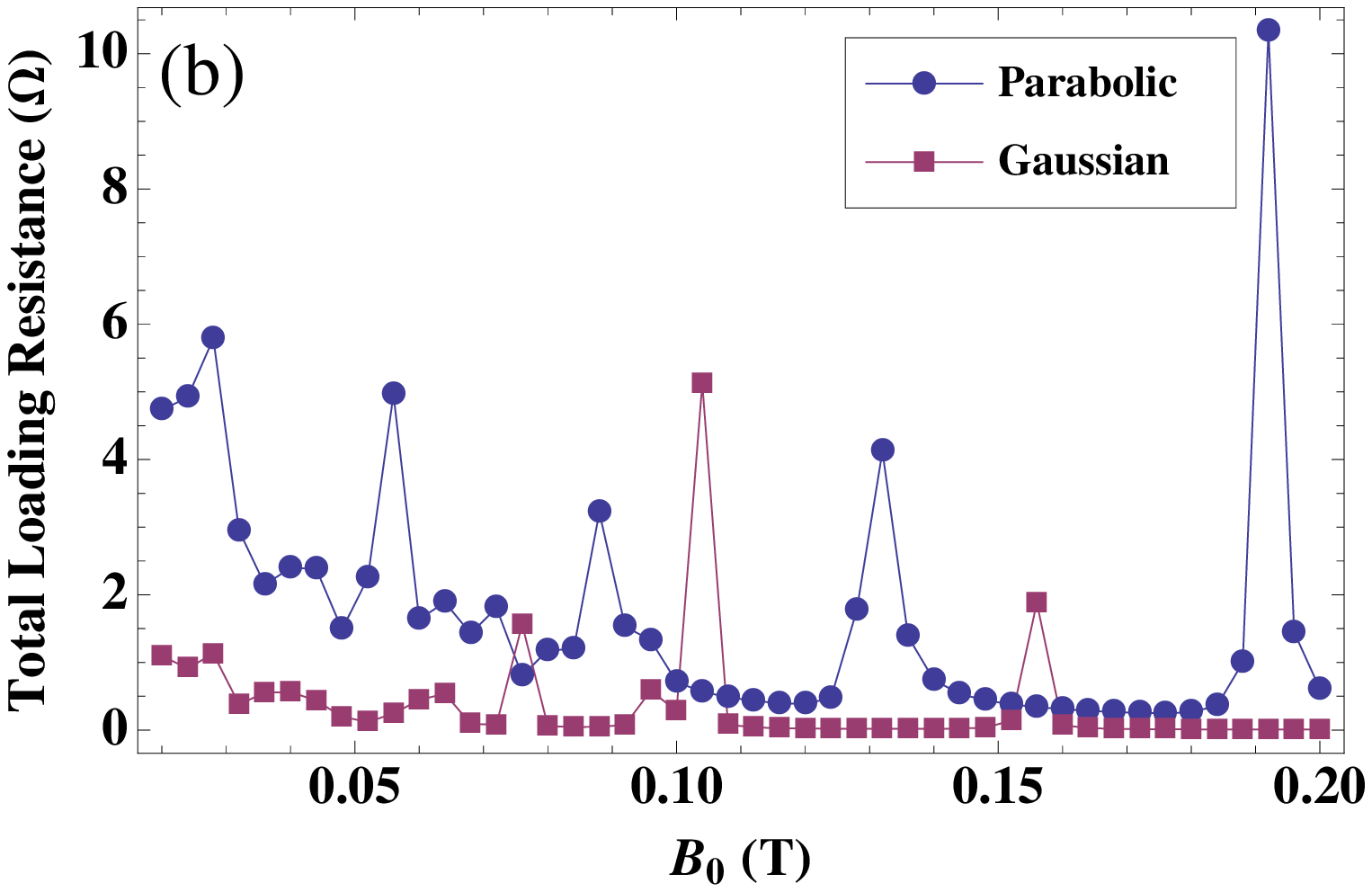}\\
\includegraphics[width=0.505\textwidth,angle=0]{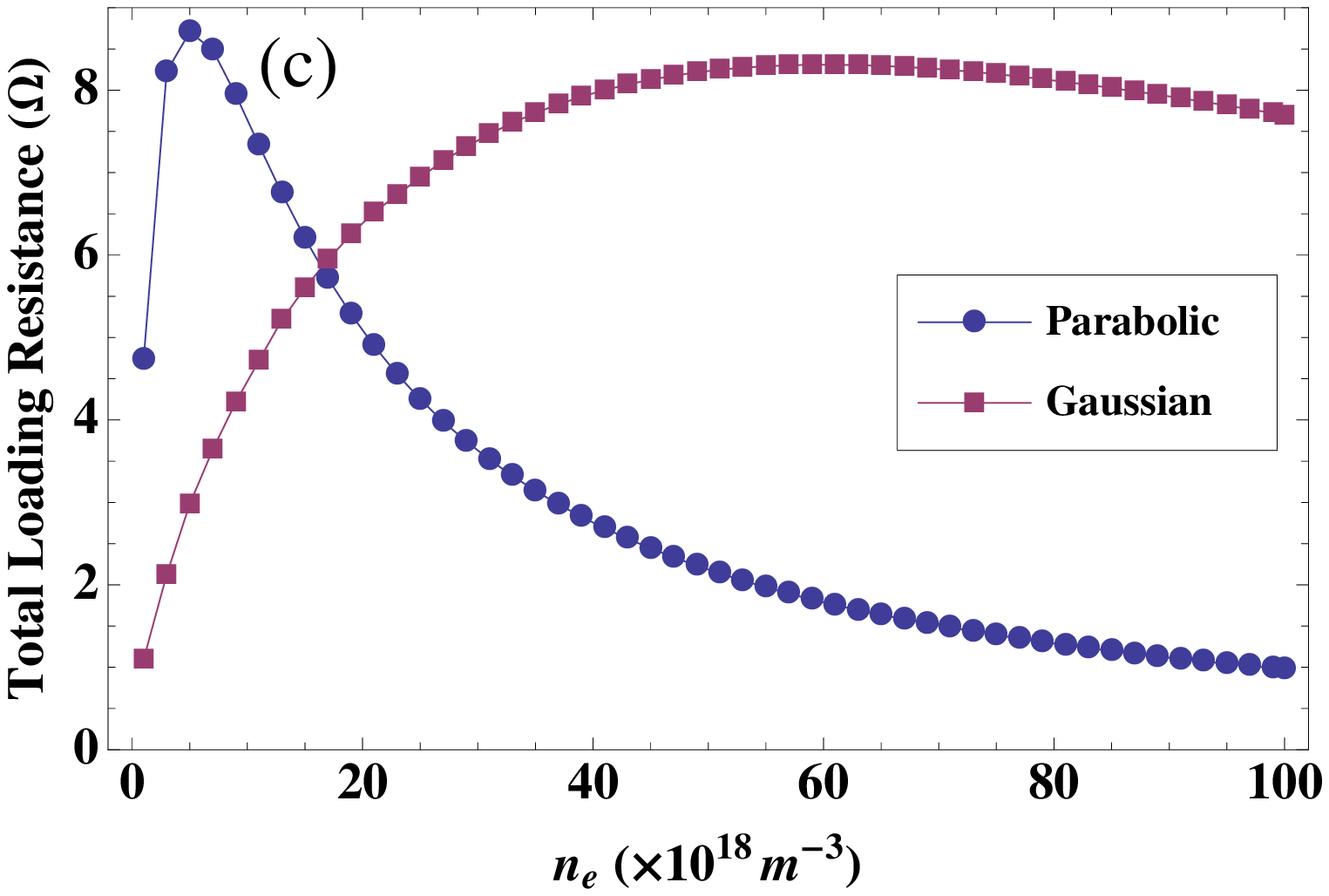}
\end{array}$
\end{center}
\caption{Variations of total loading resistance with: (a) driving frequency, (b) static magnetic field strength, (c) plasma density.}
\label{fg8}
\end{figure}

To further study the coupling of RF antennas to large volume helicon plasma, we compute the total loading resistance, which directly reflects the coupling level between driving antenna and plasma column, by scanning driving frequency, static magnetic field strength ($B_0$) and plasma density on axis ($n_0$). The scanning procedure starts from $f=13.56$~MHz, $B_0=0.02$~T and $n_0=10^{18}~\mbox{m}^{-3}$, by varying one parameter at each time. As shown in Fig.~\ref{fg8} (a), higher frequency results in larger total loading resistance, which indicates better coupling and more power absorption. This conclusion agrees well with Sec.~\ref{scan} and previous studies\cite{Chen:2012aa, Gushchin:2012aa}. The dependence of total loading resistance on field strength is given in Fig.~\ref{fg8} (b), from which we can see multiple local peaks labeling maximum power absorption at certain values of $B_0$. But overall the coupling decreases with field strength and is consistent with a previous work\cite{Gushchin:2012aa}. \textcolor{blue}{These local peaks may be associated with the resonance between field strength and wave number, which happens when the relation of $3.83/R_p=\omega e \mu_0 n_e/(k B_0)$ ($R_p$ is plasma radius)\cite{Chen:1996ab} is satisfied for discrete $k$.} Again, the parabolic density profile leads to higher power absorption than Gaussian density profile, as shown in Fig.~\ref{fg8} (a-b) and previous sections. However, this overwhelming superiority reverses when the density increases above a certain value. As shown in Fig.~\ref{fg8} (c), the total loading resistance for parabolic density profile drops quickly and is well below that of Gaussian density profile when the density increases over $n_0\approx18\times10^{18}~\mbox{m}^{-3}$. Therefore, for low-density plasma, parabolic density profile is better than Gaussian density profile in terms of coupling efficiency, whereas for high-density plasma the latter is better. The variation profile of power absorption with Gaussian plasma density resembles the result obtained by others\cite{Soltani:2016aa}. 

\section{Summary}
Motivated by enhancing the coupling of RF antennas to large volume helicon plasma for high power applications, we study the dependences of power absorption on antenna geometry and driving frequency for helicon source with infinite length and diameter of $0.5$~m. Theoretical model based Maxwell's equations and cold-plasma dielectric tensor is employed together with the associated well-known $\textbf{HELIC}$ code, and two typical density profiles in radius (parabolic and Gaussian) are considered. We find that, compared with the $m=1$ mode driven antennas such as half helix, Boswell and Nagoya III, the $m=0$ mode driven loop antenna produces overall higher power absorption in both radial and axial directions. The absorption is axially symmetric about the antenna location except the half helix antenna, which has axial preference and is related to the directions of antenna twist and static magnetic field. Higher driving frequency leads to higher power absorption, and moves the absorption location radially more outward. The parabolic density profile results in more broader absorption spectrum in $k$ than Gaussian density profile, for which the absorption mainly happens around small value of $k$. The perpendicular stream pattern of wave electric field changes significantly when the driving frequency is increased, whereas the patterns of wave magnetic field and perturbed current remain nearly unchanged. Finally, to see the whole plasma response and its parameter dependence, we compute the total loading resistance for various driving frequencies, strengths of static magnetic field and plasma densities. It is found that the total loading resistance, which reflects the coupling and matching level between antenna and plasma, increases with the driving frequency; there exist multiple local peaks for certain field strengths, indicating the maximum absorption there, but overall the total loading resistance decreases with field strength; more importantly, the parabolic density profile does not always attract more power from antenna than Gaussian density profile but only for low density situations, whereas for high density the latter attracts more power absorption. These conclusions are particular interesting and useful for the experimental design of high power helicon sources which usually have large diameters. 

\ack
Inspiring discussions with Yuankai Peng and Xinjun Zhang are appreciated. This work is supported by the National Natural Science Foundation of China (11405271), China Postdoctoral Science Foundation (2017M612901), Chongqing Science and Technology Commission (cstc2017jcyjAX0047), Chongqing Postdoctoral Special Foundation (Xm2017109), Fundamental Research Funds for Central Universities (20822041A4261), and Pre-research of Key Laboratory Fund for Equipment (61422070306). 

\section*{References}
\bibliographystyle{unsrt}

\begin{thebibliography}{10}

\bibitem{Lieberman:2005aa}
M.~A. Lieberman and A.~J. Lichtenberg.
\newblock {\em Principles of Plasma Discharges and Materials Processing}.
\newblock John Wiley \& Sons, Inc., Hoboken, New Jersey, second edition
  edition, 2005.

\bibitem{Diaz:2000aa}
F.~R. Chang-D{\'\i}az.
\newblock The vasimr rocket.
\newblock {\em Scientific American}, 283(5):90--97, 2000.

\bibitem{Diaz:2001aa}
F.~R. Chang-D{\'\i}az.
\newblock An overview of the vasimr engine: high power space propulsion with rf
  plasma generation and heating.
\newblock In {\em 14th Topical Conference on Radio Frequency Power in Plasmas},
  volume 595, pages 3--15, Oxnard, USA, 2001.

\bibitem{Shinohara:2013aa}
S.~Shinohara, T.~Tanikawa, T.~Hada, I.~Funaki, H.~Nishida, T.~Matsuoka,
  F.~Otsuka, K.~P. Shamrai, T.~S. Rudenko, and T.~Nakamura.
\newblock High-density helicon plasma sources: basics and application to
  electrodeless electric propulsion.
\newblock In {\em Joint Conference of 9th International Conference on Open
  Magnetic Systems for Plasma Confinement (OS) and 3rd International Workshop
  on Plasma Material Interaction Facilities for Fusion Research (PMIF)},
  volume~63, pages 164--167, Tsukuba, Japan, 2013.

\bibitem{Shinohara:2014aa}
S.~Shinohara, H.~Nishida, T.~Tanikawa, T.~Hada, I.~Funaki, and K.~P. Shamrai.
\newblock Development of electrodeless plasma thrusters with high-density
  helicon plasma sources.
\newblock {\em IEEE Transactions on Plasma Science Transactions on Plasma
  Science}, 42(5):1245--1254, 2014.

\bibitem{Rapp:2016aa}
J.~Rapp, T.~M. Biewer, T.~S. Bigelow, J.~B.~O. Caughman, R.~C. Duckworth, R.~J.
  Ellis, D.~R. Giuliano, R.~H. Goulding, D.~L. Hillis, R.~H. Howard, T.~L.
  Lessard, J.~D. Lore, A.~Lumsdaine, E.~J. Martin, W.~D. McGinnis, S.~J.
  Meitner, L.~W. Owen, H.~B. Ray, G.~C. Shaw, and V.~K. Varma.
\newblock The development of the material plasma exposure experiment.
\newblock {\em IEEE Transactions on Plasma Science}, 44(12):3456--3464, Dec
  2016.

\bibitem{Rapp:2017aa}
J.~Rapp, T.~M. Biewer, T.~S. Bigelow, J.~F. Caneses, J.~B.~O. Caughman, S.~J.
  Diem, R.~H. Goulding, R.~C. Isler, A.~Lumsdaine, C.~J. Beers, T.~Bjorholm,
  C.~Bradley, J.~M. Canik, D.~Donovan, R.~C. Duckworth, R.~J. Ellis, V.~Graves,
  D.~Giuliano, D.~L. Green, D.~L. Hillis, R.~H. Howard, N.~Kafle, Y.~Katoh,
  A.~Lasa, T.~Lessard, E.~H. Martin, S.~J. Meitner, G.~N. Luo, W.~D. McGinnis,
  L.~W. Owen, H.~B. Ray, G.~C. Shaw, M.~Showers, V.~Varma, and the MPEX~team.
\newblock Developing the science and technology for the material plasma
  exposure experiment.
\newblock {\em Nuclear Fusion}, 57(11):116001, 2017.

\bibitem{Goulding:2017aa}
R.~H. Goulding, J.~B.~O. Caughman, J.~Rapp, T.~M. Biewer, T.~S. Bigelow, I.~H.
  Campbell, J.~F. Caneses, D.~Donovan, N.~Kafle, E.~H. Martin, H.~B. Ray, G.~C.
  Shaw, and M.~A. Showers.
\newblock Progress in the development of a high power helicon plasma source for
  the materials plasma exposure experiment.
\newblock {\em Fusion Science and Technology}, 72(4):588--594, 2017.

\bibitem{Blackwell:2012aa}
B.~D. Blackwell, J.~F. Caneses, C.~M. Samuell, J.~Wach, J.~Howard, and C.~Corr.
\newblock Design and characterization of the magnetized plasma interaction
  experiment (magpie): a new source for plasma--material interaction studies.
\newblock {\em Plasma Sources Science and Technology}, 21(5):055033, 2012.

\bibitem{Shinohara:2004aa}
S.~Shinohara and T.~Tanikawa.
\newblock Development of very large helicon plasma source.
\newblock {\em Review of Scientific Instruments}, 75(6):1941--1946, 2004.

\bibitem{Tanikawa:2006aa}
T.~Tanikawa and S.~Shinohara.
\newblock Plasma performance in very large helicon device.
\newblock {\em Thin Solid Films}, 506-507:559 -- 563, 2006.

\bibitem{Chen:1997aa}
F.~F. Chen and R.~W. Boswell.
\newblock Helicons-the past decade.
\newblock {\em Plasma Science, IEEE Transactions on}, 25(6):1245, 1997.

\bibitem{Arnush:1997aa}
D.~Arnush and F.~F. Chen.
\newblock Generalized theory of helicon waves. ii. excitation and absorption.
\newblock {\em Physics of Plasmas}, 5(5):1239, 1997.

\bibitem{Arnush:2000aa}
D.~Arnush.
\newblock The role of trivelpiece--gould waves in antenna coupling to helicon
  waves.
\newblock {\em Physics of Plasmas}, 7(7):3042, 2000.

\bibitem{Stix:1957aa}
T.~H. Stix.
\newblock Oscillations of a cylindrical plasma.
\newblock {\em Phys. Rev.}, 106:1146--1150, 1957.

\bibitem{Stix:1992aa}
T.~H. Stix.
\newblock {\em Waves in Plasmas}.
\newblock American Institute of Physics, New York, 1992.

\bibitem{Boswell:1987aa}
R.~W. Boswell and R.~K. Porteous.
\newblock Large volume, high density rf inductively coupled plasma.
\newblock {\em Applied Physics Letters}, 50(17):1130, 1987.

\bibitem{Squire:2006aa}
J.~P. Squire, F.~R. Chang-D{\'\i}az, T.~W. Glover, V.~T. Jacobson, G.~E.
  McCaskill, D.~S. Winter, F.~W. Baity, M.~D. Carter, and R.~H. Goulding.
\newblock High power light gas helicon plasma source for vasimr.
\newblock {\em Thin Solid Films}, 506-507(Supplement C):579 -- 582, 2006.

\bibitem{Mori:2004aa}
Y.~Mori, H.~Nakashima, F.~W. Baity, R.~H. Goulding, M.~D. Carter, and D.~O.
  Sparks.
\newblock High density hydrogen helicon plasma in a non-uniform magnetic field.
\newblock {\em Plasma Sources Science and Technology}, 13(3):424, 2004.

\bibitem{Shamrai:1998aa}
K.~P. Shamrai.
\newblock Stable modes and abrupt density jumps in a helicon plasma source.
\newblock {\em Plasma Sources Science and Technology}, 7(4):499--511, 1998.

\bibitem{Boswell:1970aa}
R.~W. Boswell.
\newblock Plasma production using a standing helicon wave.
\newblock {\em Physics Letters A}, 33(7):457, 1970.

\bibitem{Boswell:1984ab}
R.~W. Boswell.
\newblock Very efficient plasma generation by whistler waves near the lower
  hybrid frequency.
\newblock {\em Plasma Physics and Controlled Fusion}, 26(10):1147, 1984.

\bibitem{Boswell:1997aa}
R.~W. Boswell and F.~F. Chen.
\newblock Helicons-the early years.
\newblock {\em Plasma Science, IEEE Transactions on}, 25(6):1229, 1997.

\bibitem{Chen:1997ab}
F.~F. Chen and D.~Arnush.
\newblock Generalized theory of helicon waves. i. normal modes.
\newblock {\em Physics of Plasmas}, 4(9):3411, 1997.

\bibitem{Chang:2011aa}
L.~Chang, M.~J. Hole, and C.~S. Corr.
\newblock A flowing plasma model to describe drift waves in a cylindrical
  helicon discharge.
\newblock {\em Physics of Plasmas}, 18(4):042106, 2011.

\bibitem{Chen:1996aa}
F.~F Chen, I.~D. Sudit, and M.~Light.
\newblock Downstream physics of the helicon discharge.
\newblock {\em Plasma Sources Science and Technology}, 5(2):173, 1996.

\bibitem{Chen:1996ab}
F.~F. Chen.
\newblock Physics of helicon discharges.
\newblock {\em Physics of Plasmas}, 3(5):1783, 1996.

\bibitem{Sudit:1996aa}
I.~D. Sudit and F.~F Chen.
\newblock Discharge equilibrium of a helicon plasma.
\newblock {\em Plasma Sources Science and Technology}, 5(1):43, 1996.

\bibitem{Lee:2011aa}
C.~A. Lee, G.~Chen, A.~V. Arefiev, R.~D. Bengtson, and B.~N. Breizman.
\newblock Measurements and modeling of radio frequency field structures in a
  helicon plasma.
\newblock {\em Physics of Plasmas}, 18(1):013501, 2011.

\bibitem{Chang:2012aa}
L.~Chang, M.~J. Hole, J.~F. Caneses, G.~Chen, B.~D. Blackwell, and C.~S. Corr.
\newblock Wave modeling in a cylindrical non-uniform helicon discharge.
\newblock {\em Physics of Plasmas}, 19(8):083511, 2012.

\bibitem{Soltani:2016aa}
B.~Soltani, M.~Habibi, and H.~Zakeri-khatir.
\newblock The effect of radial inhomogeneity on the collisional power
  absorption in helicon plasma sources.
\newblock {\em Physics of Plasmas}, 23(2):023507, 2016.

\bibitem{Chen:2012aa}
F.~F. Chen.
\newblock Performance of a permanent-magnet helicon source at 27 and 13?mhz.
\newblock {\em Physics of Plasmas}, 19(9):093509, 2012.

\bibitem{Gushchin:2012aa}
M.~E. Gushchin, T.~M. Zaboronkova, C.~Krafft, S.~V. Korobkov, and A.~V.
  Kostrov.
\newblock Inductance and near fields of a loop antenna in a cold magnetoplasma
  in the whistler frequency band.
\newblock {\em Physics of Plasmas}, 19(9):093301, 2012.

\end{thebibliography}

\end{document}